\documentclass[prd,aps,showpacs,preprintnumbers,amssymb]{revtex4}
\usepackage{graphicx}% Include figure files

\usepackage{epsf}

\def\e3p{$\eta \rightarrow 3 \pi$}

\begin{document}

\title{%
\hfill{\normalsize\vbox{%
\hbox{\rm SU-4252-889}
 }}\\
{ Gauged linear sigma model and pion-pion
scattering}}

\author{Amir H. Fariborz,
$^{\it \bf a}$~\footnote[1]{Email:
 fariboa@sunyit.edu}}

\author{N.W. Park
$^{\it \bf b,c}$~\footnote[2]{Email:
   nwpark@jnu.ac.kr          }}

\author{Joseph Schechter
 $^{\it \bf b}$~\footnote[4]{Email:
 schechte@phy.syr.edu}}

\author{M. Naeem Shahid
$^{\it \bf b}$~\footnote[5]{Email:
   mnshahid@phy.syr.edu            }}

\affiliation{$^ {\bf \it b}$ Department of Physics,
 Syracuse University, Syracuse, NY 13244-1130, USA,}
\affiliation{$^ {\bf \it a}$ Department of Mathematics/
Science, State University of New York, Institute of Technology,Utica, NY13504-3050, USA,}
\affiliation{$^ {\bf \it c}$ Department of Physics,
 Chonnam National University, Gwangju, 500-757, South Korea,}
\date{\today}

\begin{abstract}
 A simple gauged linear sigma model with
several parameters to take
 the symmetry breaking and the mass differences
 between the vector meson and the axial vector
 meson into account is considered here as a
possibly useful ``template" for the role of a light scalar
in QCD as well as for (at a different scale)
 an effective Higgs sector for some recently proposed
walking technicolor models. An analytic procedure is
first developed for relating the Lagrangian parameters to
four well established (in the QCD application) experimental
inputs. One simple equation distinguishes three 
different cases:i. QCD with axial vector particle 
heavier than vector particle, ii. possible technicolor
model with vector particle heavier than the axial vector one,
iii. the unphysical QCD case where both the KSRF and 
Weinberg relations hold. The model is applied to the 
s-wave pion-pion scattering in QCD. Both the near 
threshold region and (with an assumed unitarization)
the ``global" region up to about 800 MeV are considered.
 It is noted that there is a little tension between the choice
 of ``bare" sigma mass parameter for describing these 
 two regions. If a reasonable ``global" fit is made, there
 is some loss of precision in the near threshold region.

\end{abstract}


\maketitle

\section{introduction}

    The linear sigma model \cite{gl} containing the pion
field and a scalar field, sigma has played a 
very
 important role
in particle physics during the last half century.
Originally, it helped to clarify the role of chiral
symmetry and its spontaneous breakdown \cite{njl}
 in strong interaction 
physics. The partial conservation of the axial
 currents together
with their algebraic structure resulted in calculations
for low energy pion physics which gave, for the first
 time, reasonable 
agreements with
experiment.  While the original calculations \cite{ad} were
roundabout, it was found that they could be greatly
simplified by straightforward perturbative calculations in 
the non-linear version of the model obtained by assuming the
sigma field to be very heavy. Experimental evidence at that
time did not clearly demand a light sigma meson.

    Remarkably, the original linear version also
turned out to be useful at a higher energy scale as the 
Higgs potential \cite{ws} of the electroweak standard model, with 
the sigma identified as the Higgs field.

   More recently there has been some renewal of interest in the
linear model for two different
reasons. First, many people began to accept that a light sigma  
plays an important, but slightly hidden, role in low energy pion
scattering. Second, the idea of ``walking technicolor" theories
\cite{wt}
has gained some popularity. These theories furnish 
interesting candidates
for an {\it effective} Higgs sector which has some similarity to
the effective chiral Lagrangians associated with the color QCD
theory. Such effective Lagrangians usually include,
in addition to spin 0 fields, spin 1 fields
since ``vector meson dominance" is a well established feature
of low energy strong interactions. Especially, the interplay 
of the composite technicolor vector and axial vector bosons
appears to play an important role.  

    Our initial motivation is the further understanding of the
properties of light scalar mesons. This is essentially
connected with the s-wave pion pion scattering problem and 
also with the generalization to three light flavors of
the underlying quarks.
Some characteristic papers in the recent revival of
interest in this subject are \cite{kyotoconf}-\cite{b}.
A fascinating aspect is the
possibility that the light scalars contain two quarks
and two antiquarks rather than one quark and one antiquark
\cite{j}. Consideration of the surprisingly light masses of 
such a ``four quark" scalar nonet together with the surprisingly
heavy candidates for a different, conventional ``two quark"
nonet
suggests \cite{BFS3} a mixing between the two nonets. As
discussed in section V of \cite{BFMNS01}, 
it seems instructive to formulate this in a chiral invariant
way using a generalized SU(3) linear sigma model. This enables
one to formally distinguish four quark vs two quark mesons
by means of their differing axial U(1) transformations as well
as to compare four quark contents of scalar and
 pseudoscalar states. Other related work on mixing includes
\cite{mixing}- \cite{ythb}. Now in these generalized
sigma models
 there are more than one scalar and the s-wave scattering is more 
complicated. At least qualitatively, the s- channel is dominated
by these scalars. In the region near threshold,
 the lowest mass sigma is most important. Thus, as a start to
studying the effects of the vector mesons in such models, it
seems natural to go back to the original linear sigma model
 and add the 
vector meson, rho with its chiral partner. Of course, this
is not a new subject. A review of older work is given in 
\cite{gg} and recent papers include those in 
\cite{recentsigmod} . 

    Even though the plain linear sigma model is quite
 simple to deal
with, the addition of spin 1 fields increases the overall
complexity by an order of magnitude.
    In this paper
the  vector and axial vector fields are added
as (initially) Yang Mills gauge fields. The local gauge
symmetry is then manifestly broken by the addition of the 
three simplest chiral invariant spin 1 field mass terms.
A somewhat new feature of the present work is that
the determination of all Lagrangian parameters 
is carried out analytically 
with respect to the experimental inputs. The s-wave pion
scattering is studied both for the threshold region 
and for the region 
away from threshold which is expected to be influenced
by the presence of the lightest sigma. The scattering is
explicitly compared with that of the plain
 linear sigma model as well as with 
experiment.  

    Because we have a simple connection between
 the Lagrangian
parameters and the physical inputs, it is straightforward
for us to also discuss application of the model
to the non-QCD, but presumably chiral, situation which describes
the walking technicolor theories. In particular, the case
 \cite{fran}
where the vector meson is heavier than the axial vector 
meson will be treated in detail. 

In section II, we expand the model
 Lagrangian in terms of the component
fields representing the scalar $\sigma$, the pseudoscalar,
${\vec \pi}$ the vector ${\vec V}_\mu$ and the axial vector,
${\vec A}_\mu$. Section III first discusses
 the diagonalization necessitated by the ${\vec A}_\mu
\cdot
\partial_\mu{\vec \pi}$ term in the Lagrangian. This results
in ``physical" pion and axial vector fields denoted with tildes.
In general, the tilde will also be used on other objects
to denote the fact that they are ``physical".
Especially, in this section, the
crucial job of determining the parameters of the Lagrangian
from experiment will be discussed in detail.
In section IV the complicated formula for the invariant
Mandelstam amplitude in pion-pion scattering is
computed at tree level. The s-wave partial wave
amplitudes with I=0 and I=2 are then given explicitly.
Section V discusses these amplitudes near threshold
and also briefly explains their unitarizations
by the K-matrix method. In section VI, the behaviors
of the s-wave amplitudes away from threshold are treated.
Section VII discusses some connections of this model with
some interesting other work; first the limit of the present model
in which the Weinberg \cite{Weinberg} and
Kawarabayashi Suzuki Riazuddin Fayazuddin (KSRF) \cite{KSRF}
 relations both hold is 
discussed.
In addition, a systematic treatment of the model is given for the
case in which (going beyond the application to low energy QCD)
the vector meson mass is greater than the axial vector meson mass.
Section VIII provides additional discussion. In the Appendix,
it is explicitly shown how the scattering amplitude
reproduces the ``current algebra" 
 result of the chiral model without 
spin 1 fields
in the limit where the sigma mass goes to infinity.

\section{Lagrangian}

Here, we present the version of the $SU_L(2)\times SU_R(2)$
 gauged linear sigma model Lagrangian
to be studied.
%we adopt to study the scattering amplitude.
The basic fields are the scalar, $\sigma$
and pion, $\vec{\pi}$, which are contained in
%collectively expressed by
 $M=(\sigma+i\vec{\pi}\cdot\vec{\tau})/\sqrt{2}$
 and its Hermitian conjugate. The
starting piece is the kinetic term for $M$, which is
invariant under the chiral transformation,
$M^\prime = U_L M U_R^{-1}$. One
can naturally introduce
the left ($l_\mu$) and the right ($r_\mu$)
 vector fields by gauging the chiral symmetry.
  The resulting gauge
invariant Lagrangian density is then,

\begin{eqnarray}\label{Lag}
{\cal L}
=-\frac{1}{2} Tr(F_{\mu\nu}^rF_{\mu\nu}^r
+F_{\mu\nu}^lF_{\mu\nu}^l)
 -\frac{1}{2} Tr(D_{\mu}M^\dag D_{\mu}M),
\end{eqnarray}

where the covariant derivatives of $M$ and $M^\dagger$ are,

\begin{eqnarray}\label{covar}
D_{\mu}M & = & \partial_\mu M-igl_\mu M+igMr_\mu,\nonumber \\
D_{\mu}M^\dag & = & \partial_\mu M^\dag-igr_\mu M^\dag+igM^\dag
l_\mu ,
\end{eqnarray}

 and the field strength tensors take the form,

\begin{eqnarray}\label{field}
F_{\mu\nu}^l & = & \partial_\mu l_\nu-\partial_\nu
l_\mu-ig[l_\mu,l_\nu],
 \nonumber \\
F_{\mu\nu}^r & = & \partial_\mu r_\nu-\partial_\nu
r_\mu-ig[r_\mu,r_\nu].
\end{eqnarray}

The vector and axial vector mesons are defined as
\begin{eqnarray}\label{defva}
V_{\mu}&=&l_\mu+r_\mu=\frac{1}{\sqrt{2}}
\vec{V}_\mu\cdot\vec{\tau},\nonumber\\
A_{\mu}&=&l_\mu-r_\mu=\frac{1}{\sqrt{2}}
\vec{A}_\mu\cdot\vec{\tau}.
\end{eqnarray}

%Since all mesons have masses, we need mass terms which
% break gauge invariance and chiral invariance.
The terms which contribute to particle masses are:

\begin{eqnarray}\label{Lag1}
- m_0^2Tr(l_{\mu}l_{\mu}+r_{\mu}r_{\mu})
+B Tr(Mr_{\mu}M^\dag l_{\mu})
-C Tr(l_{\mu}^2 M M^\dag +r_{\mu}^2 M^\dag M)
- V_0(M,M^\dag)-V_{SB}.
\end{eqnarray}

The first, $m_0^2$ term, which breaks the gauge invariance
(and also the formal scale
symmetry), gives the same mass to the vector
 and the axial vector mesons.
The C term also gives the same mass to both spin 1 mesons,
 but maintains the scale symmetry.
 The B term breaks the mass degeneracy of the two
 spin 1 mesons. This is important since, experimentally,
 the lightest isovector, axial vector meson with negative
G-parity (the $a_1$(1260) is heavier than the rho meson.
Another contribution to this mass splitting arises
 from spontaneous chiral symmetry breaking in the model, but
this effect by itself will be seen to be insufficient.
  The last two terms
 are the scalar potential terms which respectively
 yield the spontaneous chiral symmetry breaking
 and the explicit symmetry breaking due
 to the small quark mass; explicitly,

\begin{equation}\label{vsb}
V_0(M,M^\dag)=a_1(\sigma^2+
{\vec \pi}\cdot{\vec \pi})+a_3(\sigma^2+
{\vec \pi}\cdot{\vec \pi}))^2,
\hspace{1cm}
V_{SB}=-2\sqrt{2}A\sigma.
\end{equation}

Here, $a_3$ is positive while
 $a_1$ is chosen to be negative so that spontaneous
chiral symmetry breaking
 will give a nontrivial
 vacuum expectation value $v$ for $\sigma$.
 The explicit symmetry breaking term $V_{SB}$
mocks up the light quark mass terms.
 The coefficients in this potential can be determined
 by the minimum condition in terms of the
 sigma and pion mass parameters, with the definition 
 $V\equiv V_0+V_{SB}$, as follows:

\begin{eqnarray}\label{Lag1}
<\frac{\partial V}{\partial\sigma}>=0=
2a_1v+4a_3v^3-2\sqrt2 A,
\nonumber\\
<\frac{\partial^2 V_0}{\partial\sigma^2}>=
m_{\sigma}^2=2a_1+12a_3v^2, \nonumber\\
<\frac{\partial^2 V_0}{\partial\pi^2}>=
m_\pi^2=2a_1+4a_3 v^2.
\end{eqnarray}

From this, one can easily derive the coefficients,

\begin{eqnarray}\label{vsb2}
m_\pi^2&=&\frac{2\sqrt{2}A}{v},\nonumber\\
a_1&=&\frac{1}{2}(m_\sigma^2-\frac{3}{2}
(m_\sigma^2-m_\pi^2)),\nonumber\\
a_3&=&\frac{m_\sigma^2-m_\pi^2}{8v^2}.
\end{eqnarray}

The potential terms can be expressed
in terms of the fields as:

\begin{eqnarray}\label{vsb3}
V_0(M,M^\dag)&=&\frac{1}{2} m_\pi^2
{\vec \pi}\cdot{\vec \pi}
+\frac{1}{2}
 m_\sigma^2 \sigma^2+\frac{1}{2} g_{\sigma\pi\pi}\sigma
{\vec \pi}\cdot{\vec \pi}
+\frac{1}{4} g_4 ({\vec \pi}\cdot{\vec \pi})^2+... ,
\nonumber \\
 g_{\sigma\pi\pi}&=&\frac{m_\sigma^2-m_\pi^2}{v},
\hspace{1cm}
 g_4=\frac{2g_{\sigma\pi\pi}}{v}.
\end{eqnarray}
Here, quadrilinear terms
involving $\sigma$ have not been written.
Also note that the quantities $m_\pi$,
$g_{\sigma\pi\pi}$ and $g_4$ are not the
 physical ones, which will
be defined later.

Now, we express the rest of the
 Lagrangian in terms of the
component fields. The spin
zero meson kinetic terms are:

\begin{eqnarray}\label{spinzerokinet}
-\frac{1}{2}Tr(D_{\mu}M
D_{\mu}M^\dag)=&-&\frac{1}{2}\partial_\mu\vec{\pi}\cdot
\partial_\mu\vec{\pi}
-\frac{1}{2}\partial_\mu\sigma\partial_\mu\sigma
+\frac{g}{\sqrt{2}}\vec{A}_\mu\cdot
(\sigma\stackrel{\leftrightarrow}{\partial_\mu}\vec{\pi})
-\frac{g}{2\sqrt{2}}\epsilon_{abc}V_{\mu
a}(\pi_b\stackrel{\leftrightarrow}{\partial_\mu}\pi_c)\nonumber\\
+g^2[-\frac{\sigma^2}{4}\vec{A_\mu}\cdot\vec{A_\mu}
&+&\frac{1}{2}\epsilon_{abc}\sigma\pi_aV_{\mu b}A_{\mu c}
+\frac{1}{4}(\vec{\pi}\cdot\vec{V_\mu})^2
-\frac{1}{4}(\vec{\pi}\cdot\vec{\pi})(\vec{V_\mu}\cdot\vec{V_\mu})
-\frac{1}{4}(\vec{\pi}\cdot\vec{A_\mu})^2].
\end{eqnarray}

Here, it is understood that $\sigma=v+\tilde\sigma$
where $\tilde\sigma$ is the physical $\sigma$ field.

 The Yang-Mills terms are:
\begin{eqnarray}\label{ymterms}
-\frac{1}{2} Tr(F_{\mu\nu}^rF_{\mu\nu}^r+F_{\mu\nu}^lF_{\mu\nu}^l)
&=&-\frac{1}{4}[(\partial_\mu V_{\nu a}-\partial_\nu V_{\mu a})^2
+(\partial_\mu A_{\nu a}-\partial_\nu A_{\mu a})^2]
\nonumber \\
&-&\frac{g}{2\sqrt{2}}\epsilon_{abc}[(\partial_\mu V_{\nu c}
-\partial_\nu V_{\mu c})(V_{\mu a} V_{\nu b}+
A_{\mu a} A_{\nu b})\nonumber\\
&-&(\partial_\mu A_{\nu c} -\partial_\nu A_{\mu c})(V_{\mu a} A_{\nu
b}+A_{\mu a} V_{\nu b})] \nonumber \\
&-&\frac{g^2}{8}[(\vec{V_\mu}\cdot\vec{V_\mu})^2
-(\vec{V_\mu}\cdot\vec{V_\nu})^2
+(\vec{A_\mu}\cdot\vec{A_\mu})^2-
(\vec{A_\mu}\cdot\vec{A_\nu})^2\nonumber\\
&+&2(\vec{V_\mu}\cdot\vec{V_\mu})(\vec{A_\nu}\cdot\vec{A_\nu})
-2(\vec{V_\mu}\cdot\vec{V_\nu})(\vec{A_\mu}\cdot\vec{A_\nu})
\nonumber \\
&+&4(\vec{V_\mu}\cdot\vec{A_\mu})(\vec{V_\nu}\cdot\vec{A_\nu})
-2(\vec{V_\mu}\cdot\vec{A_\nu})(\vec{V_\mu}\cdot\vec{A_\nu})].
\end{eqnarray}

Finally, the spin one meson mass terms are:

\begin{eqnarray}\label{spinonemtype}
-m_0^2 Tr(l_{\mu}l_{\mu}+r_{\mu}r_{\mu})
&=&-\frac{1}{2}m_0^2(\vec{V_\mu}\cdot
\vec{V_\mu}+\vec{A_\mu}\cdot\vec{A_\mu}),\nonumber\\
-CTr(l_{\mu}^2 M M^\dag +r_{\mu}^2 
M^\dag M)&=&-\frac{C}{4}(\vec{V_\mu}\cdot\vec{V_\mu}+
\vec{A_\mu}\cdot\vec{A_\mu})(\sigma^2+
{\vec \pi}\cdot{\vec \pi}),\nonumber\\
BTr(Mr_{\mu}M^\dag l_{\mu})&=&B[\frac{1}{8}\sigma^2(\vec{V_\mu}
\cdot\vec{V_\mu}-\vec{A_\mu}\cdot\vec{A_\mu})
-\frac{1}{2}\epsilon_{abc}\sigma\pi_aV_{\mu b}A_{\mu c}
+\frac{1}{4}(\vec{\pi}\cdot\vec{V_\mu})^2\nonumber\\
&&-\frac{1}{8}(\vec{\pi}\cdot\vec{\pi})
(\vec{V_\mu}\cdot\vec{V_\mu})
-\frac{1}{4}(\vec{\pi}\cdot\vec{A_\mu})^2
+\frac{1}{8}(\vec{\pi}\cdot\vec{\pi})
(\vec{A_\mu}\cdot\vec{A_\mu})].
\end{eqnarray}

\section{Diagonalization and Determination of Parameters}

In this model, we have the five parameters
$g$, $v$, $m_0^2$, B and C
 to be determined from
experiment. $g$ and $v$ are intrinsic parameters of the model
while $m_0^2$, B and C represent different ways to introduce
vector and axial vector masses. Specifically the vector and
axial vector masses are given by:
\begin{eqnarray}\label{spinonemass}
m_V^2&=&m_0^2-\frac{Bv^2}{4}+C\frac{v^2}{2},\nonumber\\
m_A^2&=&m_0^2+\frac{Bv^2}{4}+C\frac{v^2}{2}
+\frac{g^2v^2}{2}
\equiv {m^\prime}_0^2+\frac{g^2v^2}{2}.
\end{eqnarray}
We see, from the third term
on the right hand side of
Eq.(\ref{spinzerokinet}), that
 the Lagrangian yields a
  pion-axial vector meson mixing
 term  proportional to $v \vec{A}_\mu \cdot
\partial_\mu\vec{\pi}$.
The Lagrangian can be diagonalized
 by introducing the physical (tilde) quantities as

\begin{eqnarray}\label{diagonal}
\vec{A}_\mu&=&\tilde{\vec{A}_\mu}+
b\partial_\mu\tilde{\vec{\pi}}, \nonumber\\
\vec{\pi}&=&w\tilde{\vec{\pi}}.
\end{eqnarray}

b is determined from the condition of
 zero mixing between the physical pion
and the physical axial vector meson, while
 w is determined from the condition of correct
normalization of the pion
 kinetic term. We find
\begin{eqnarray}\label{renor1}
b&=&\frac{gv}{\sqrt{2} w {m^\prime}_0^2},\nonumber\\
w&=&\sqrt{1+\frac{g^2v^2}{2{m^\prime}_0^2}}.
\end{eqnarray}
The following alternate forms also are useful:
\begin{eqnarray}\label{renor}
b=\frac{gvw}{\sqrt{2} m_A^2}\nonumber\\
w^2=\frac{m_A^2}{{m^\prime}_0^2}=
\frac{1}{1-\frac{g^2v^2}{2m_A^2}}
\end{eqnarray}
Note that ${m^\prime}_0^2$ was defined in
Eq.(\ref{spinonemass}) above.
The physical pion decay constant,
 $\tilde{F_\pi}$ is obtained from the
Noether's theorem calculation of the
single particle contributions to the
 axial current
in our Lagrangian:
  \begin{eqnarray}\label{renor2}
(J_\mu^A)_1^2&=&-\sqrt2 v \Big(\frac{\partial L}
{\partial(\partial_\mu \pi)}\Big)_1^2
=\partial_\mu (\frac{{\vec \tau}
\cdot{\vec \pi}}{\sqrt2})_1^2 -
\frac{gv}{\sqrt2}(\frac{{\vec \tau}\cdot {\vec A}_\mu}
{\sqrt2})_1^2 \nonumber\\
&=&\frac{\sqrt2 v}{w} \partial_\mu
\tilde{\pi}^{+}-g v^2 {\tilde A}_\mu^+ ,
\end{eqnarray}
where we used Eq.(\ref{diagonal}) and, for example,
$\tilde{\pi}^{+}$ is the physical positive pion field.

The coefficient in front of
$\partial_\mu{\tilde \pi^+}$
 is identified as the physical pion decay constant:
\begin{eqnarray}\label{renor2}
\tilde{F_\pi}=\frac{\sqrt2 v}{w}.
\end{eqnarray}

 Finally, we determine the $\rho\pi\pi$
interaction terms. For this purpose we
 collect the terms of this type
from Eqs.(\ref{spinzerokinet}) and (\ref{ymterms}) wherein
 the pion- axial vector diagonalization given in
Eq.(\ref{diagonal}) is taken into account:

\begin{eqnarray}\label{grhopipi}
{\cal L}_{\rho\pi\pi}=\epsilon_{abc}V_{\mu
a}\tilde{\pi}_b\partial_\mu\tilde{\pi}_c
[-\frac{g}{\sqrt{2}}w^2+\frac{wg^2bv}{2}+\frac{wBbv}{2}]
-\frac{gb^2}{\sqrt{2}}\epsilon_{abc}\partial_\mu V_{\nu
c}\partial_\mu\tilde{\pi}_a\partial_\nu\tilde{\pi}_b.
\end{eqnarray}

The last term constitutes a three derivative piece
of the $\rho\pi\pi$ interaction. Note that, if one
wishes to express it as an effective single derivative
interaction, there will be a different result for
$\rho\rightarrow\pi\pi$ decay (where $\rho$ is on mass
shell) and $\pi$-$\pi$ scattering (where $\rho$ is off mass
shell).

% we can calculate corresponding Feynman diagram.
% Here, $q_1+q_2=p$, we find
%\begin{eqnarray}\label{grhopipi}
%\frac{gb^2}{\sqrt{2}}\epsilon_\sigma [m_\pi^2(-iq_{2\sigma})
%\epsilon_{fgc}+m_\pi^2(-iq_{1\sigma})\epsilon_{fgc}
%+\epsilon_{fgc}(-iq_1)\cdot(-iq_2)(-iq_{2\sigma})
%+\epsilon_{gfc}(-iq_1)\cdot(-iq_2)(-iq_{1\sigma})]\nonumber\\
%=-i\frac{gb^2}{\sqrt{2}}\epsilon_{fgc}
%[m_\pi^2\epsilon\cdot(q_2-q_1)-(q_1\cdot q_2)
%\epsilon\cdot(q_2-q_1)]\nonumber\\
%=-i\frac{gb^2}{\sqrt{2}}\epsilon_{fgc}
%(m_\pi^2-q_1\cdot q_2)\epsilon\cdot(q_2-q_1)]\nonumber\\
%=-i\frac{gb^2}{\sqrt{2}}\epsilon_{fgc}
%\frac{m_\rho^2}{2}\epsilon\cdot(q_2-q_1)
%\end{eqnarray}
% where we used
%$q_1\cdot q_2=\frac{1}{2}p^2-\frac{1}{2}<q_1>^2-
%\frac{1}{2}<q_2>^2=\frac{1}{2}(-m_\rho^2+2m_\pi^2)$
%in last line.

 From this, we can obtain the effective $\rho\pi\pi$
coupling constant for on-shell rho as:
\begin{eqnarray}\label{grhopipi}
g^{eff}_{\rho\pi\pi}=g(1-\frac{Bv^2}{2{m^\prime}_0^2}-
\frac{b^2}{2}m_\rho^2).
\end{eqnarray}

The coupling constant $g^{eff}_{\rho\pi\pi}$
 is related to the $\rho$ meson width by
\begin{eqnarray}\label{equav2}
\Gamma(\rho)=(g^{eff}_{\rho\pi\pi})^2
 | q_\pi |^3 /(12\pi m_\rho^2).
\end{eqnarray}

For $\Gamma(\rho)=149.4$ MeV,
 one finds $|g^{eff}_{\rho\pi\pi}|\approx 8.66$.

 Now we will solve for the vacuum value, $v$ by
the following procedure. First replace $w$ in
the second of
 Eqs.(\ref{renor1}) by, from Eq.(\ref{renor2}),
the quantity $\sqrt{2}v/{\tilde F_\pi}$.
Then replace $2{m^\prime}_0^2$ by, using
Eqs.(\ref{spinonemass}), $2m_A^2-g^2v^2$.
Squaring both sides gives the quadratic equation for $v^2$:
\begin{eqnarray}\label{equav2}
v^4-\frac{2m_A^2}{g^2} v^2 +
\frac{2m_A^2 {\tilde F_\pi}^2}{2g^2} =0.
\end{eqnarray}
This can be solved easily
 in terms of $g^2 v^2$ to get:
\begin{eqnarray}\label{2signs}
g^2v^2=m_A^2\Big(1\pm\sqrt{1-\frac{g^2
 {\tilde F}_\pi^2}{m_A^2}}\,\,\Big).
\end{eqnarray}

This is an equation which determines the product
$gv$ in terms of g and experimentally known quantities.
We can find another relation between $g$ and $v$
by substituting $Bv^2/2=m_A^2-m^2_\rho -g^2v^2/2$
and $b=gvw/(\sqrt{2}m_A^2)$ into Eq.(\ref{grhopipi}):
\begin{eqnarray}\label{geff}
g^{eff}_{\rho\pi\pi} =g\bigg(1-\frac{1}{2m_A^2 -g^2v^2}
\Big(2(m_A^2-m_\rho^2)-g^2v^2(1-
\frac{m_\rho^2}{2m_A^2})\Big)\bigg )
\end{eqnarray}
Substituting Eq.(\ref{2signs}) into Eq.(\ref{geff}) gives
an equation for the Yang-Mills coupling constant, $g$
by itself. Knowing this we can substitute back into
Eq.(\ref{2signs})to determine v. Then we can determine
B from:
\begin{equation}
 B=\frac{2}{v^2}(m_A^2-m^2_\rho) -g^2.
\label{findB}
\end{equation}
Finally, we may determine the linear combination,
$m_0^2+Cv^2/2$ from:
\begin{equation}
m_0^2+Cv^2/2=(m_\rho^2+m_A^2)/2 -g^2v^2/4.
\label{findmandrho}
\end{equation}
Note that from the four given inputs it is
only possible to obtain the given linear combination
of $m_0^2$ and C. Later we will consider two different
``models" corresponding to either $m_0$ =0 or $m_0 \ne$ 0.
Table \ref{params} shows the results based on the
best fit value of $m_A$ as well as its maximum and minimum
values. Note also that the solution requires the sign
in Eq.(\ref{2signs}) to be positive.
The solutions with zero value for the square root
 and with the minus sign will be discussed in
a later section.

\begin{table}[htbp]
\begin{center}
\begin{tabular}{c|c|c|c|c|c|c}
\hline
$m_A$ in GeV  & g   & v in GeV  &w & b in $GeV^{-1}$& B & $m_0^2$ +
 $Cv^2/2$ in GeV$^2$ \\
\hline \hline
  1.270 & 7.83 & 0.2&2.2& 1.55&-12.9& 0.456\\
 1.230  & 7.78& 0.197& 2.13 &1.53& -13.73& 0.467 \\
 1.190  & 7.72& 0.19& 2.06 & 1.51&-14.65& 0.468 \\
\hline
\end{tabular}
\end{center}
\caption[]{
 $g,v,w,b,B,m_0^2+Cv^2/2$ as functions of the axial vector
 meson mass.
We used ${\tilde F}_\pi$=0.131 GeV,
 $m_\rho$=0.775 GeV, $g_{\rho\pi\pi}^{eff}$=8.56
as inputs.
Note that g, w and B are dimensionless.}
\label{params}
\end{table}

It can be seen that the predicted parameters
 are not  much affected by the uncertainty in the mass 
of the $a_1(1260)$ meson. Thus we will use the central
 value in what follows.

\section{Pion-pion scattering amplitude}

    At tree level, the
conventional Mandelstam scattering amplitude,
A(s,t,u)
 has the following
contributions:

  1) Zero derivative contact term:
\begin{eqnarray}
&&     - {\tilde g}_{\sigma\pi\pi}w^2/v,
\nonumber \\
&&
{\tilde
g}_{\sigma\pi\pi}\equiv \frac{w^2}{v}
(m_\sigma^2-\frac{{\tilde m}_\pi^2}{w^2}).
   \label{amp1}
\end{eqnarray}
Note that Eqs.(\ref{vsb3}) and (\ref{diagonal})
 were used in
obtaining this result.

    2) Two derivative contact term:
\begin{equation}\label{2dercon}
(\frac{g^2}{2}+B-C)b^2w^2 s-
Bb^2 {\tilde m}_\pi^2w^2+2b^2C {\tilde m}_\pi^2w^2.
\end{equation}
Note that the factor $b^2$
is due to the presence of a physical
pion field in the original
axial vector meson field, $A_\mu$, as described in
the first of Eqs.(\ref{diagonal}). Thus, $b^2$
labels the two derivative interaction terms.

   3) Four derivative contact term:
\begin{eqnarray}\label{4dercon}
-\frac{g^2}{2}b^4(2s^2-t^2-u^2-12{\tilde m}_\pi^2s
+16{\tilde m}_{\pi}^4)
\end{eqnarray}
Note, as above, that the $b^4$ factor indicates
 these terms arise from the quartic
Yang-Mills interaction
 of the axial vector gauge field.

   4) Sigma pole in the s- channel:
\begin{equation}
\label{sigamp}
\frac{1}{m_\sigma^2-s}[-{\tilde g}_{\sigma\pi\pi}
+\sqrt{2}{\tilde m}_\pi^2gbw-2G({\tilde m}_\pi^2-
\frac{s}{2})]^2,
\end{equation}
where,
\begin{equation}
 G=-\frac{vg^2b^2}{2}-\frac{vBb^2}
{4}+\frac{2gbw}{\sqrt2}
-\frac{C}{2}b^2v.
\label{G}
\end{equation}

    5) Rho poles in the t and u channels:
\begin{equation}
\label{rhoamp}
\frac{s-u}{m_\rho^2-t}[-G_1+\frac{gb^2}{2\sqrt{2}}
t]^2+\frac{s-t}{m_\rho^2-u}[-G_1+
\frac{gb^2}{2\sqrt{2}}u]^2,
\end{equation}
where,
\begin{equation}
\label{G1}
G_1=\frac{g}{\sqrt{2}}(1-\frac{Bv^2}{2{m^\prime}_0^2}).
\end{equation}

   The full amplitude A(s,t,u) is, of course, the
sum of all five pieces just written. Since it is
rather complicated, we verify in the Appendix,
that in the $m_\sigma$ goes to infinity limit
A(s,t,u) reduces to the correct "current algebra
form.
Here, we will be interested in the
  $I=0,2$ projections:
\begin{eqnarray}
T^0&=&3A(s,t,u)+A(t,u,s)+A(u,s,t),\nonumber\\
T^2&=&A(t,u,s)+A(u,s,t),
\label{isospin}
\end{eqnarray}
where the Mandelstam variables
 are $s=4(p_\pi^2+{\tilde m}_\pi^2)$,
 $t=-2p_\pi^2(1-cos\theta)$, $u=-2p_\pi^2(1+cos\theta)$,
$p_\pi$ being the spatial momentum of the pion
in the center of mass frame.

The angular momentum $l$ partial
 wave elastic scattering amplitude for isospin $I$ is
then defined as,
\begin{equation}
T_{l}^I=\frac{1}{64\pi}\sqrt{1-\frac
{4{\tilde m}_\pi^2}{s}}\int_{-1}^1 dcos\theta
P_{l}(cos\theta)T^I(s,t,u).
\label{angmom}
\end{equation}

Using the above formula, we get:

\begin{eqnarray}\label{t00}
T_0^0&=&\frac{1}{64\pi}\sqrt{1-
\frac{4{\tilde m}_\pi^2}{s}}\, \Big[
10(\frac{m_\sigma^2 -{\tilde m}_\pi^2/w^2}{v^2}w^4
+(2C-B)b^2w^2{\tilde m}_\pi^2)+
\frac{6}{m_\sigma^2-s}[-{\tilde
g}_{\sigma\pi\pi}
+\sqrt{2}{\tilde m}_\pi^2gbw-2G({\tilde m}_\pi^2-
\frac{s}{2})]^2\nonumber\\
&&+2(C_1^2 S_1+2C_1 G S_2+G^2 S_3)
+4(G_1^2 R_1+G_1 \frac{gb^2}{\sqrt{2}}
 R_2+ \frac{g^2b^4}{8} R_3)-
\frac{3g^2b^4}{8} (4s^2-
\frac{64p_\pi^4}{3}-24{\tilde m}_\pi^2 s
+32{\tilde m}_\pi^4)\nonumber\\
&&-\frac{g^2b^4}{4}(-2s^2+\frac{32p_\pi^4}{3}
+48 {\tilde m}_\pi^2 s+32{\tilde m}_\pi^4)+6 b^2w^2(\frac{g^2}{2}+
B-C)s+2b^2w^2(\frac{g^2}{2}+B-C)(-4p_\pi^2)\Big]
\end{eqnarray}

where
\begin{eqnarray}
\label{CSR}
&& C_1= -{\tilde g}_{\sigma\pi\pi}+
\sqrt{2}{\tilde m}_\pi^2gbw-2G {\tilde m}_\pi^2,
\nonumber \\
&&S_1=\frac{1}{2p_\pi^2}
 ln(\frac{m_{\sigma}^2+4p_\pi^2}{m_\sigma^2}),
S_2=m_\sigma^2 S_1-2, S_3=4 p_\pi^2+m_\sigma^2 S_2,
\nonumber \\
&&R_1=\frac{1}{2p_\pi^2}
ln(\frac{m_\rho^2 +4p_\pi^2}{m_\rho^2})
(s+m_\rho^2 +4p_\pi^2)-2,
 R_2=m_\rho^2 R_1-4p_\pi^2-2s,R_3=
m_\rho^2 R_2+\frac{16p_\pi^4}{3}+4p_\pi^2 s.
\end{eqnarray}
 Similarly for the $I=2$ case:

\begin{eqnarray}\label{t02}
T_0^2&=&\frac{1}{64\pi}\sqrt{1-\frac{4{\tilde m}_\pi^2}{s}}\,
\Big[
4(\frac{m_\sigma^2 -{\tilde m}_\pi^2/w^2}{v^2}
w^4+(2C-B)b^2w^2{\tilde
m}_\pi^2)
\nonumber\\
&&-4(C_1^2 S_1+2C_1 G S_2+G^2 S_3)
+4(G_1^2 R_1+2G_1 gb R_2+ gb^2 R_3)\nonumber\\
&&-\frac{g^2b^4}{4}(-2s^2+\frac{32p_\pi^4}{3}+48 {\tilde m}_\pi^2
 s+32m_\pi^4)+2b^2w^2(\frac{g^2}{2}+B-C)(-4p_\pi^2)\Big]
\end{eqnarray}

\section{Scattering near threshold}
    Using the well known
experimental results for the rho mass and width as well
as  the $a_1$(1260) mass, we specified in Table \ref{params}
 the Lagrangian parameters $g$, $v$, $w$, $b$, $B$ and
the linear combination
$m_0^2+ Cv^2/2$. The only remaining ``unknowns" 
are the ``bare" mass of the sigma, $m_\sigma$ and the
relative sizes of $m_0^2$ and $C$.
For definiteness we will initially consider the case,
 $m_0$ =0; soon
we will see that the case, $m_0\ne$ 0, gives a poorer
fit in the region away from threshold.
Then the near threshold scattering will depend just on 
the value, $m_\sigma$. Of course one first considers the
s-wave scattering lengths. 

\begin{figure}[h]
\begin{center}
\vskip 1cm
\epsfxsize = 7.5cm
\ \epsfbox{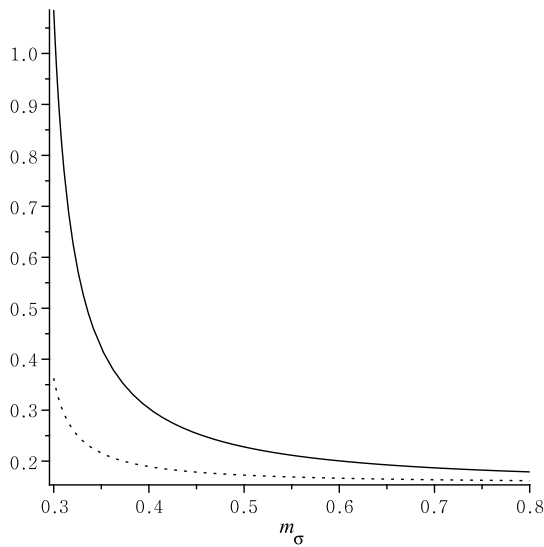}
\end{center}
\caption[]{%
The scattering length $m_\pi a_0^0$ as a function of
the sigma mass in GeV. The solid line: pure linear sigma model.
 The dotted line: the present model including spin 1 mesons.}
\label{vplot1}
\end{figure}

\begin{figure}[h]
\begin{center}
\vskip 1cm
\epsfxsize = 7.5cm
\ \epsfbox{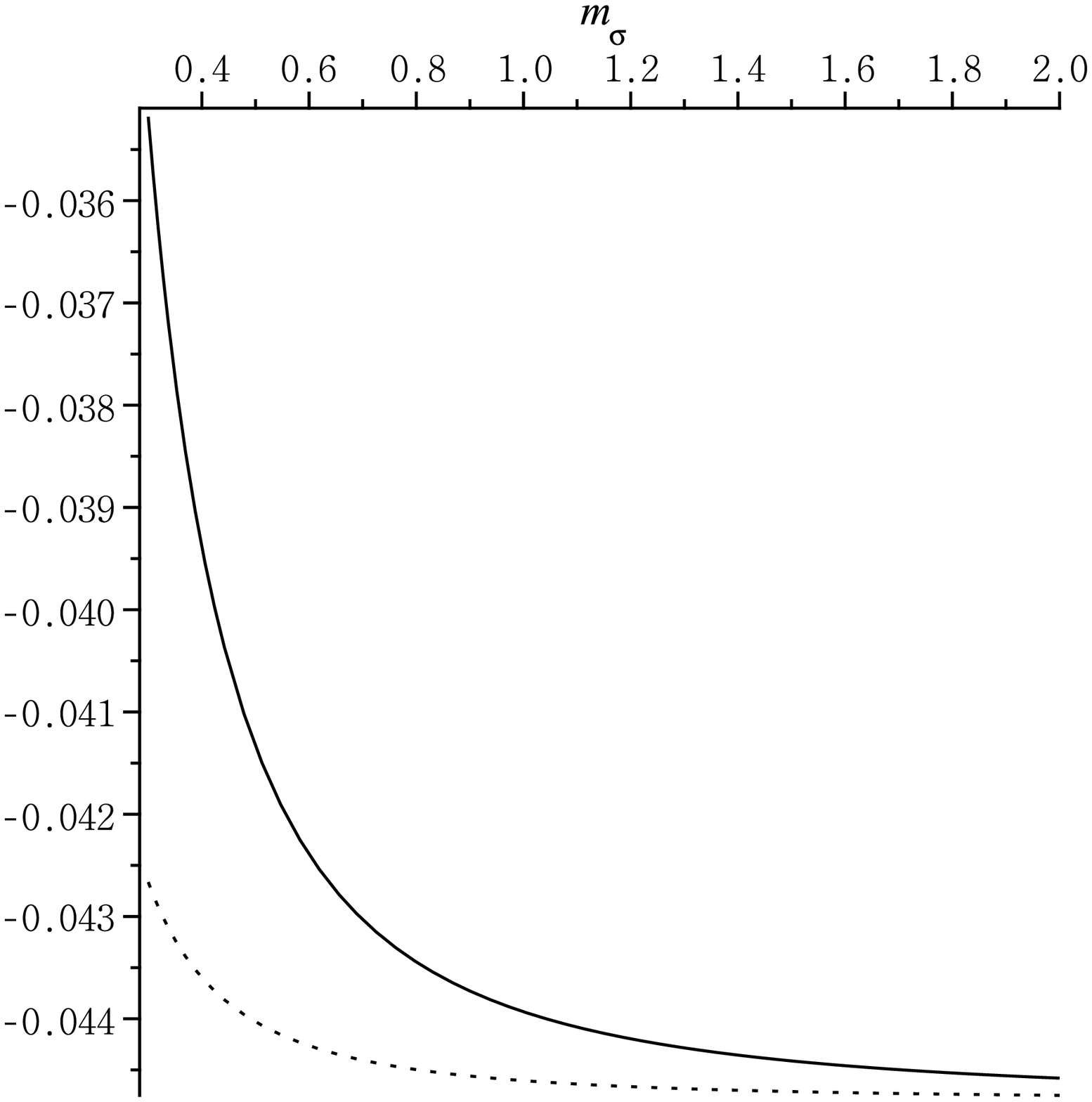}
\end{center}
\caption[]{%
 The scattering length $m_\pi a_2^0$ as a function of
the sigma mass in GeV. The solid line: pure linear
 sigma model. The dotted line:present model including
spin 1 mesons.}
\label{vplot2}
\end{figure}

    The scattering length $m_\pi a_0^0$ is
 plotted in Fig. \ref{vplot1} as a function of $m_\sigma$. 
 Also shown are the predictions in the case of
the ``pure" linear sigma model, in which the vector and axial
vector mesons are absent. It is seen that any given value
of $m_\pi a_0^0$ (above the ``current algebra" value
of about 0.16 \cite{car})
may be obtained for some $m_\sigma$. However,
for a given value of the scattering length, 
 $m_\sigma$ is seen to be substantially lower when the
 vector and axial
vector mesons are present. The experimental value of about
0.22 is obtained for $m_\sigma\approx$ 550 MeV in the plain
linear sigma model but for $m_\sigma\approx$ 360 MeV in 
the model containing the spin 1 mesons.
 Fig.\ref{vplot2} similarly shows
 the dependence of the non-resonant
scattering length, $m_\pi a_0^2$ on $m_\sigma$.

\begin{figure}[h]
\begin{center}
\vskip 1cm
\epsfxsize = 7.5cm
\ \epsfbox{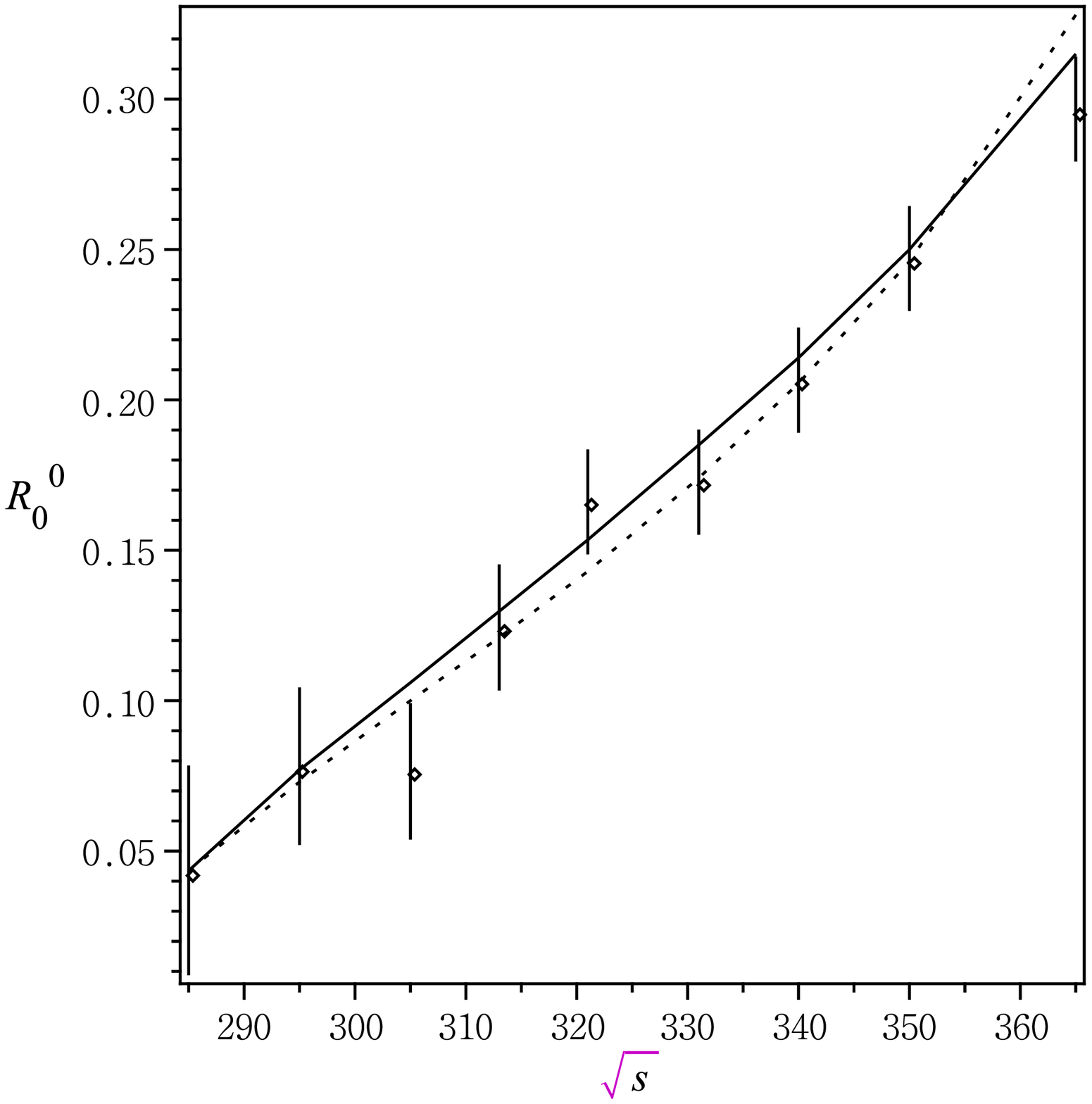}
\end{center}
\caption[]{%
Low energy data for real part of s-wave
resonant amplitude plotted against $\sqrt{s}$ in MeV. 
The dotted curve is a fit to the present
model with $m_\sigma$ = 0.42 GeV while the solid curve is a fit
to the plain linear sigma model with $m_\sigma$ = 0.62 GeV}.
\label{vplot3}
\end{figure}

    Note that, for example,
\begin{equation}
{\tilde m}_\pi a_0^0= \frac{T_0^0}{\rho},
\hspace{.5in} \rho=\sqrt{1-4{\tilde m}_\pi^2/s},
\label{momfactor}
\end{equation}
wherein $T_0^0/\rho$ is evaluated at threshold, remembering 
to first cancel the overall factor of $\rho$ in $T_0^0$.
The amplitude is purely real in the present tree
approximation. It is clearly convenient to 
compare with the real part of the partial wave amplitude.
 The experimental
 real part, $R_0^0$ is related to
the experimental
  phase shift, $\delta^0_0$ as
\begin{equation}
R_0^0=\frac{1}{2}sin(2\delta^0_0).
\label{phasedef}
\end{equation}
In Fig.\ref{vplot3}, for orientation, some values
of $R_0^0$ near threshold obtained 
from the phase shifts found by
the Na48/2 experiment \cite{Na48} are shown.
It can be seen that these data points
near threshold
 may be reasonably
explained by a value of $m_\sigma \approx$ 0.42 GeV
in the present model including spin 1 mesons but
with  the larger value $m_\sigma \approx$ 0.62 GeV in the
model without spin 1 mesons. It is hard 
to distinguish the two fits at the lower energies but
 above $\sqrt{s}\approx$ 0.35 MeV
 the two model curves
begin to diverge from each other and also to approach the
unitarity bound, $R_0^0$ =1/2. Clearly,
 the accuracy of the model
must be improved to obtain a ``global"
 description of the physics
which does not violate the unitarity bound.

     An easy way to cure this theoretical problem in
the present model is to use the well known
 K matrix unitarization. As applied to Eq.(\ref{t00}), we
identify the ``Born" term $T_0^0$ with $K$ and write for
the unitarized partial wave amplitude, $(T_0^0)_U$:
\begin{equation}
  (T_0^0)_U =\frac{T_0^0}{1-iT_0^0}.
\label{unitarized}
\end{equation}
Clearly, near threshold, where $T_0^0$ is small,
 the unitarized
amplitude is essentially the same as the non-unitarized one.
 This unitarization
is actually familiar in ordinary scattering
 since it converts a generic
simple pole into a Breit Wigner form. Diagrammatically,
it has the structure of a `` bubble sum". It is easy to 
verify
\cite{nochange} that the scattering length is unchanged from
the value obtained at tree level with 
this type of unitarization. Although the amplitude is
now exactly unitary, it is important to recognize that this
K-matrix procedure is, after all, a model.

\section{Scattering away from threshold}

\begin{figure}[h]
\begin{center}
\vskip 1cm
\epsfxsize = 7.5cm
\ \epsfbox{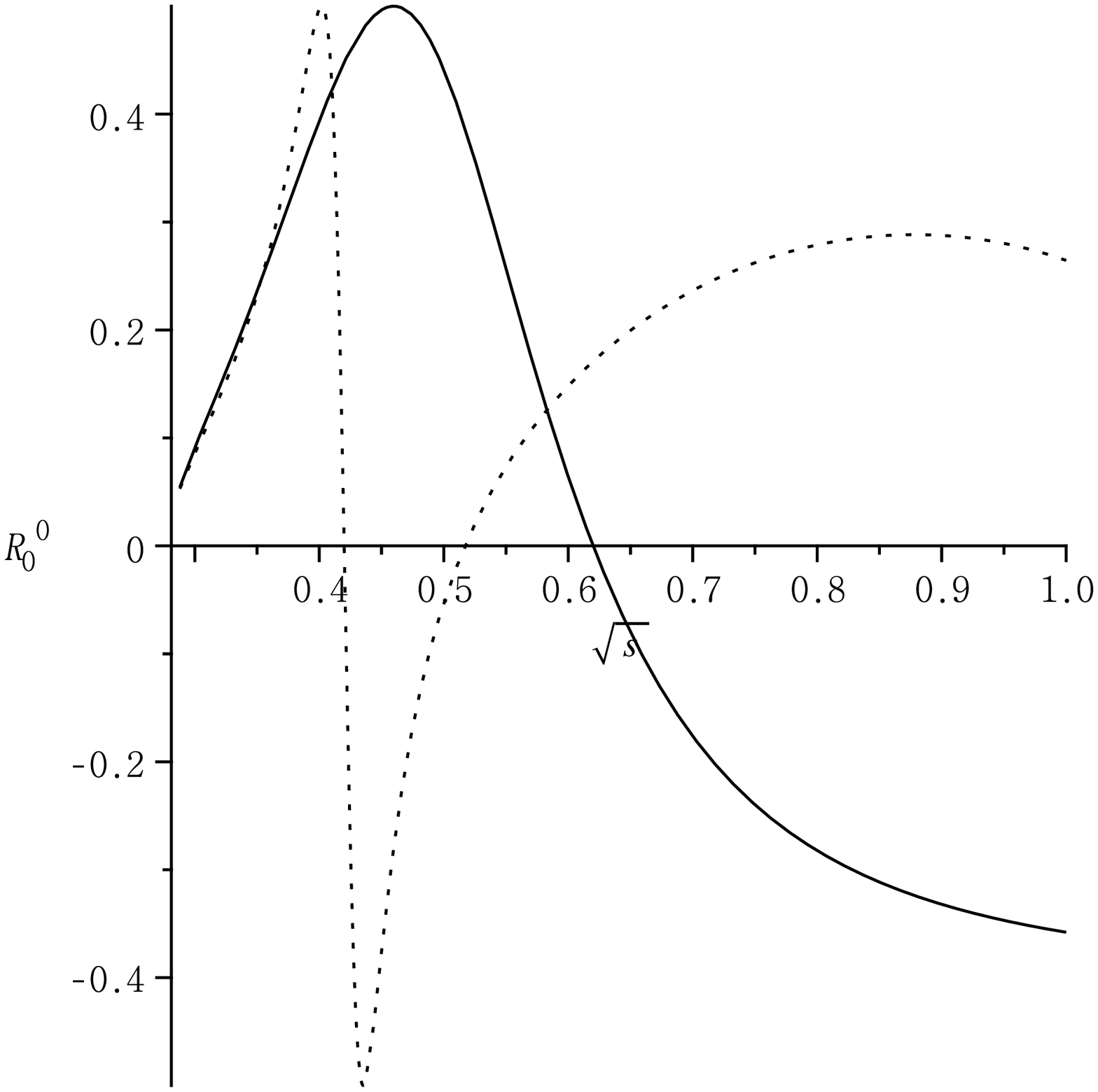}
\end{center}
\caption[]{%
 Unitarized amplitudes plotted as a function of
$\sqrt{s}$ to 1 GeV.
The dashed curve corresponds to the present model with
$m_\sigma$ = 0.42 GeV while the solid curve
corresponds to the plain linear sigma model
with $m_\sigma$ = 0.62 GeV.}
\label{vplot4}
\end{figure}

\begin{figure}[h]
\begin{center}
\vskip 1cm
\epsfxsize = 7.5cm
\ \epsfbox{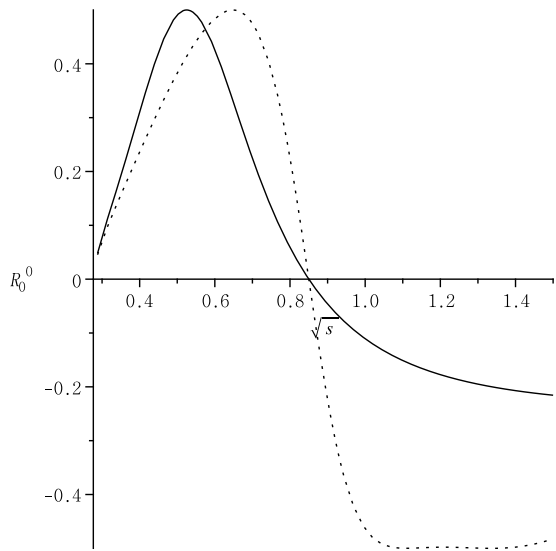}
\end{center}
\caption[]{%
Unitarized scattering amplitudes to 1.4 GeV
 with $m_\sigma$ chosen
to be 0.85 GeV for both the plain (solid curve) and 
spin 1 meson (dashed curve) sigma models.
Here $m_0^2=0$ was assumed. }
\label{vplot5}
\end{figure}

    Fig. \ref{vplot4} shows the unitarized amplitudes, just
defined, calculated up
to 1 GeV. Both the linear model with $m_\sigma$ =
0.62 GeV and the present model with additional
spin 1 fields and
 $m_\sigma$ = 0.42 GeV are again seen to start the same way.
However afterwards, the spin 1 model amplitude rises more sharply
 and has its first zero, as required [
since $R_0^0\equiv T^0_0/(1+ (T^0_0)^2)$ goes to zero when 
 $T^0_0$ goes to infinity]
 at 0.42 GeV while the
plain linear model amplitude has its first zero at
 0.62 GeV. The shapes
of these two curves do not fit  
 the experimental
data beyond the threshold region very well.
 A more realistic fit would
correspond, for example, to
  a plain linear sigma model which has its first zero
in the 0.85 GeV region; see Fig. 8 and Table II
in the first paper in ref. \cite{BFMNS01}.
(It is also seen there that the addition of the scalar
$f_0$(980) in that SU(3) linear sigma
model framework allows one to fit the peculiar
 looking amplitude from about 0.8 GeV to about 1.2 GeV.) 
As a check of the validity of this "global" fit up to
about 0.8 GeV we note that
the sigma pole position came out to be in decent
agreement with the one recently obtained by a detailed
analysis \cite{kggp} of the experimental data.
 The sigma pole position in the
complex $s$ plane is found by
separating the tree amplitude, Eq.(\ref{t00}) into pole and
 non-pole pieces as:
\begin{equation}
\label{decompose}
{T_0^0}=\alpha(s)+\frac{\beta(s)}{m^2_\sigma-s}.
\end{equation}
Then the pole position, $z$ in the complex s
plane for the K-matrix unitarized amplitude,
$(T^0_0)_U$
is the solution to the equation,
\begin{equation}
 (m_\sigma^2-z)(1-i\alpha(z))-i\beta(z)=0.
\label{polepos}
\end{equation}
 We find the numerical result in the simple K-matrix
 unitarized linear
sigma model without spin 1 particles, $z^{1/2}$ = 0.51 -i0.23. 
This may be compared with the recent value,
$z^{1/2}$ = 0.461 -i0.255, with an uncertainty of about .015
in each term.  
 In Fig. \ref{vplot5} the 
model amplitudes for both the plain linear sigma model
and the one with spin 1 particles are plotted up to
 1.4 GeV using $m_\sigma$ =0.85 GeV just mentioned.
The case including spin 1 particles was calculated
 with the choice
$m_0^2$ = 0 so that $C\ne$ 0. ( Remember
that only the combination 
$m_0^2+Cv^2/2$ is known from our inputs.)
While, as we just mentioned, the curve for the
plain linear sigma model essentially fits the data,
the curve representing the model with spin 1 particles
is a rather rough approximation to it. This can be
verified by noting that the pole position comes out to
be, $z^{1/2}$ = 0.38 -i0.52. The fit is not improved
by lowering the value of $m_\sigma$. 
 
\begin{figure}[h]
\begin{center}
\vskip 1cm
\epsfxsize = 7.5cm
\ \epsfbox{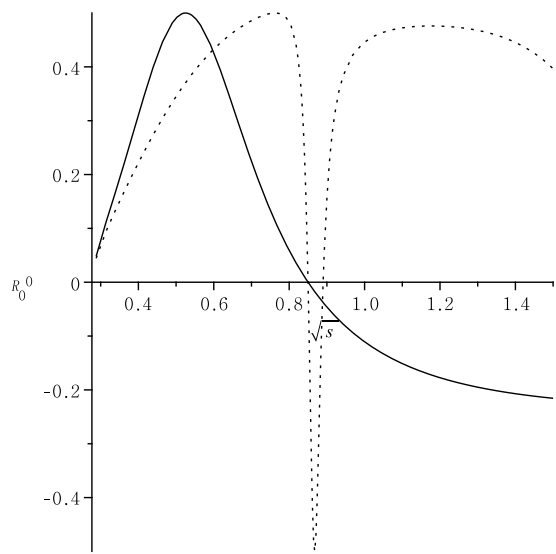}
\end{center}
\caption[]{%
Same as Fig. \ref{vplot5} but assuming $m_0^2=0.27$ 
GeV${^2}$ instead}
\label{vplot6}
\end{figure}

 It is also of some interest
to look at the dependence of the predicted amplitude
 on the parameter $m_0^2$.
for the case with spin 1 particles.
 The results for the non- zero choice,
 $m_0^2$ = 0.27 GeV$^2$ are shown in Fig. \ref{vplot6}. 
In this case the predictions for the $m_0^2\ne$ 0 case 
seem to be further distorted, showing that $m_0^2$ =0
provides a better fit.

\begin{figure}[h]
\begin{center}
\vskip 1cm
\epsfxsize = 7.5cm
\ \epsfbox{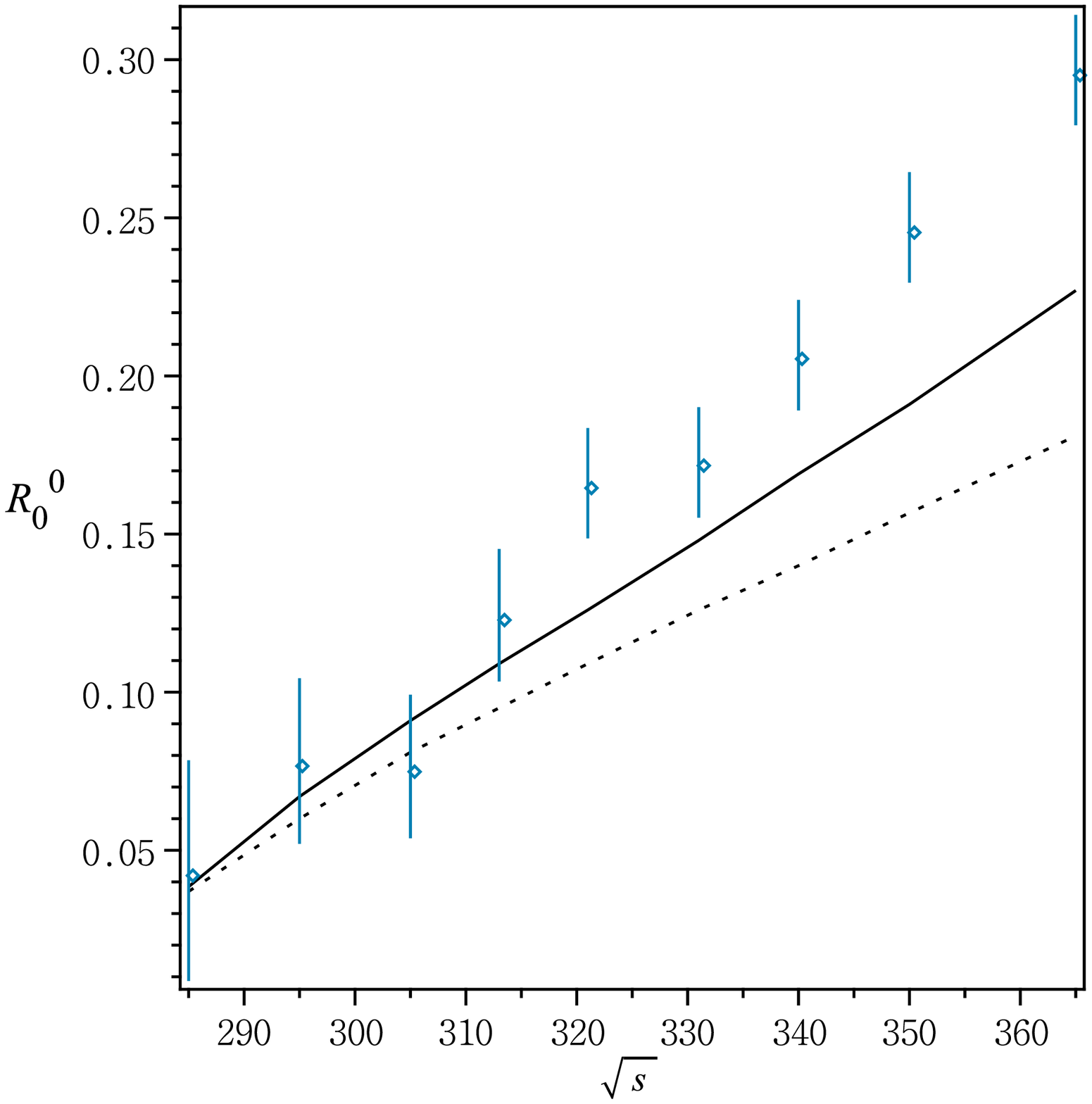}
\end{center}
\caption[]{%
Predictions for the choice $m_\sigma$ = 0.85
 GeV in the region near threshold. Same 
conventions as in Fig. \ref{vplot3} }
\label{vplot10}
\end{figure}

      How much does the $m_\sigma$ =0.85 GeV choice, which 
was used for the region up to about 0.8 GeV change
the fit to the data close to threshold obtained with smaller
values of $m_\sigma$? This is shown in Fig. \ref{vplot10}. 
Clearly, both plots lie below the low energy data.
Thus there is some tension between a reasonable fit close
to threshold (which requires a low value
of $m_\sigma$)  and a fit over a 
larger range (which requires a larger value of $m_\sigma$).

   Of course, it is clear that the direct channel
$f_0(980 MeV)$ state must be also included to adequately
 treat the scalar
I=0 amplitude in the region from 800 to about 1200 MeV.
We consider this region to be beyond the range of applicability
of the model with a single sigma state.

\section{Connections with other work}

      In the historical treatment of chiral models
containing vector and axial vector mesons as well as
the pion, two plausible relations among their
parameters - the KSRF
 \cite{KSRF} and Weinberg \cite{Weinberg}
formulas have been widely discussed. Eventually, it
was accepted that they are not forced to hold by
chiral symmetry but in some limit can be correlated
with each other.    
 These formulas are, respectively,

\begin{eqnarray}
(g^{eff}_{\rho\pi\pi})^2 &=& 2m_\rho^2/{\tilde F}_\pi^2,
\nonumber \\
m_A^2&=&2m_\rho^2.
\label{KSRFW}
\end{eqnarray}

Numerically, the first relation holds to about
 4 percent while the second only holds to
about 26 percent.
       
     In the present work it was not necessary to use
either of these formulas. Nevertheless,
 it may be interesting to first briefly discuss the
 limit of our model which 
correlates the two formulas.
This limit
 corresponds to, first, approximating
$g^{eff}_{\rho\pi\pi}$ by $g$ and, second, setting B=0.
We will show that the Weinberg  relation then implies the
KSRF relation. From both of Eqs.(\ref{spinonemass}) we then
note that $w^2$ in Eq.(\ref{renor}) becomes simply,
\begin{equation}
w^2=\frac{m_A^2}{m_\rho^2}=2.
\label{we}
\end{equation}
Eq.(\ref{renor2}) then reads $v^2={\tilde F}_\pi^2$
so that,
\begin{equation}
m_A^2-m_\rho^2=m_\rho^2= g^2v^2/2=g^2{\tilde F}_\pi^2/2,
\label{ksrfagain}
\end{equation}
which is the KSRF relation.
Note that approximating $g^{eff}_{\rho\pi\pi}$ by $g$
amounts physically to neglecting the B term
 in the Lagrangian as well as the induced
 three derivative $\rho\pi\pi$
interaction term.
It is also seen that the two equations in
Eq.(\ref{KSRFW}) hold at the special point where the
square root in Eq.(\ref{2signs}) vanishes (with B=0).

    An interesting different possible application of the
present chiral
model containing vector and axial vector mesons is to the
effective Higgs sector of the minimal walking technicolor
theory \cite{wt}. That theory may provide
the mechanism for constructing a technicolor model which
gives consistent values of the electroweak ``oblique"
parameters. A characteristic feature is
the situation where the vector boson is heavier than the axial
vector boson. To investigate this possibility we now search
for more general parameter solutions, including those
 with the negative sign in
 Eq.(\ref{2signs}).

It is convenient to define
\begin{equation}
\chi=\frac{g^2v^2}{2m_A^2}.
\label{chi}
\end{equation}
Then the pion wave function renormalization is
given by,
\begin{equation}
w^2=\frac{1}{1-\chi}.
\label{w2}
\end{equation}
Eq.(\ref{2signs}) then reads:
\begin{equation}
\chi=\frac{1}{2}\Big(1\pm\sqrt{1-\frac{g^2
 {\tilde F}_\pi^2}{m_A^2}}\,\,\Big),
\label{chig}
\end{equation}
 Notice that to have a consistent
solution for the parameters we must require:
\begin{equation}
g^2 \le m_A^2/{\tilde F}_\pi^2.
\label{gbound}
\end{equation}
Finally,  Eq.(\ref{geff}) can be
rewritten as,
\begin{equation}
g_{\rho\pi\pi}^{eff} = \frac{g\tau}{2}
(\frac{2-\chi}{1-\chi}),
\label{gsolve}
\end{equation}
where we defined, for convenience,
\begin{equation}
\tau=\frac{m_\rho^2}{m_A^2}.
\label{t}
\end{equation}
Note especially that when Eq.(\ref{chig}) is
inserted into Eq.(\ref{gsolve}), we can use it
to find $g_{\rho\pi\pi}^{eff}$ as a function of $g$
 for given values of the physical
quantities, ${\tilde F}_\pi$ and $m_A$. This 
determines $g$ and then $v$ etc.

\begin{figure}[h]
\begin{center}
\vskip 1cm
\epsfxsize = 7.5cm
\ \epsfbox{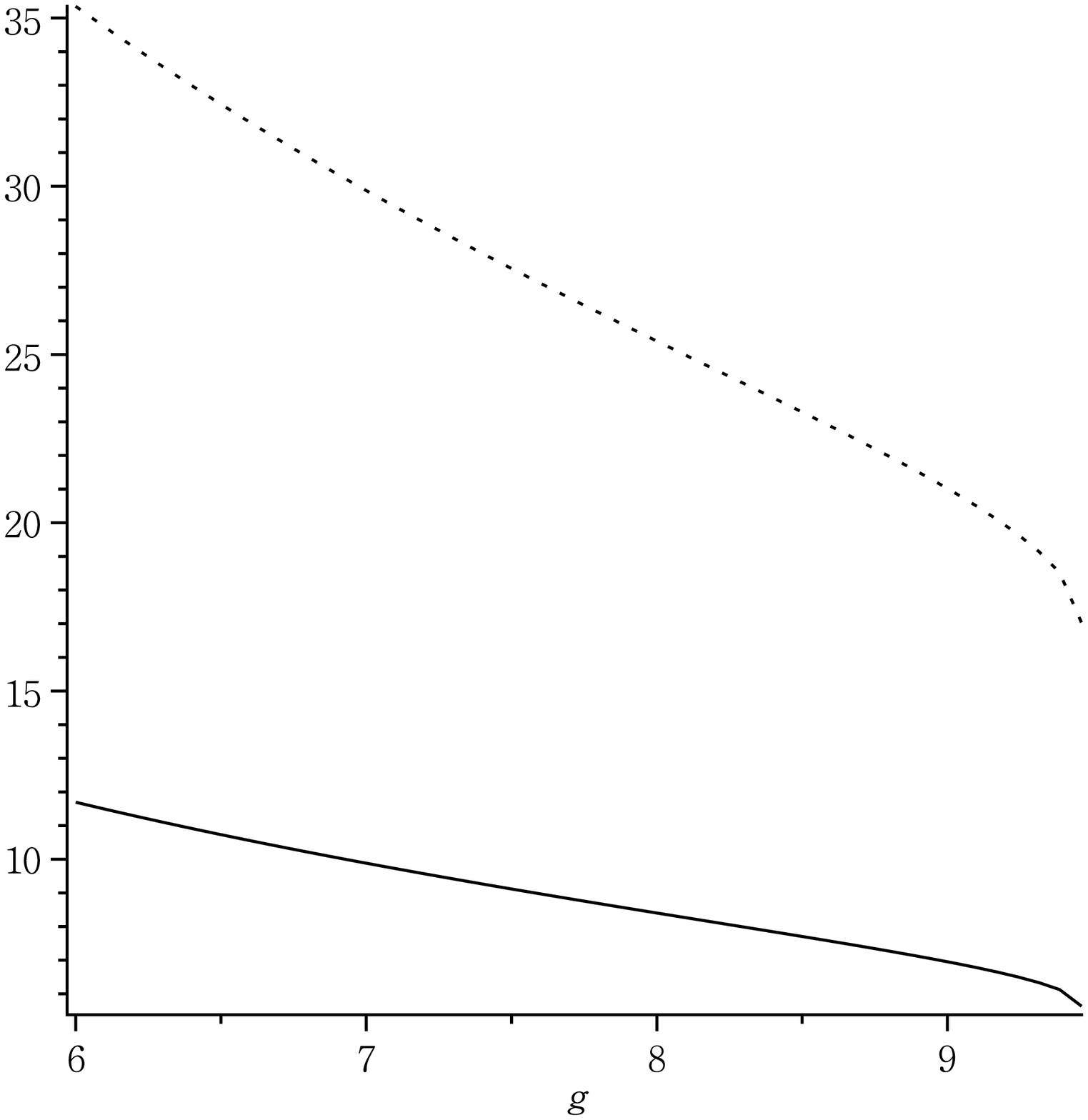}
\end{center}
\caption[]{%
$g_{\rho\pi\pi}^{eff}$ vs g with the plus
sign in Eq.(\ref{chig}). The lower curve is the
QCD case while the upper curve corresponds to
a hypothetical "walking technicolor" case 
with $m_\rho>m_A$.    
}
\label{vplot8}
\end{figure}

    In Fig.\ref{vplot8}, in which the plus sign
in Eq.(\ref{chig}) has been chosen,
 the lower curve displays
$g_{\rho\pi\pi}^{eff}$ as a function of $g$ for the physical
choice,
\begin{equation}
\tau_{QCD}=(\frac{m_{\rho}}{m_A})^2\approx 0.4.
\label{tqcd}
\end{equation}
We see that the physical value,
 $g_{\rho\pi\pi}^{eff}\approx$ 8.56 corresponds
to the value $g=$ 7.78, which is safely below
the bound at,
\begin{equation}
\frac{m_A}{{\tilde F}_\pi}=9.46.
\label{gbound}
\end{equation}
The upper curve in Fig.\ref{vplot8} corresponds,
for illustration of the $m_{\rho}>m_A$ case, to
a choice, $\tau$=1.2. In this case we have no
experimentally  a priori
way of specifying the physical parameters and the
bound. Nevertheless, we observe that
 $g_{\rho\pi\pi}^{eff}$ would be exceptionally large
for a reasonable solution.

\begin{figure}[h]
\begin{center}
\vskip 1cm
\epsfxsize = 7.5cm
\ \epsfbox{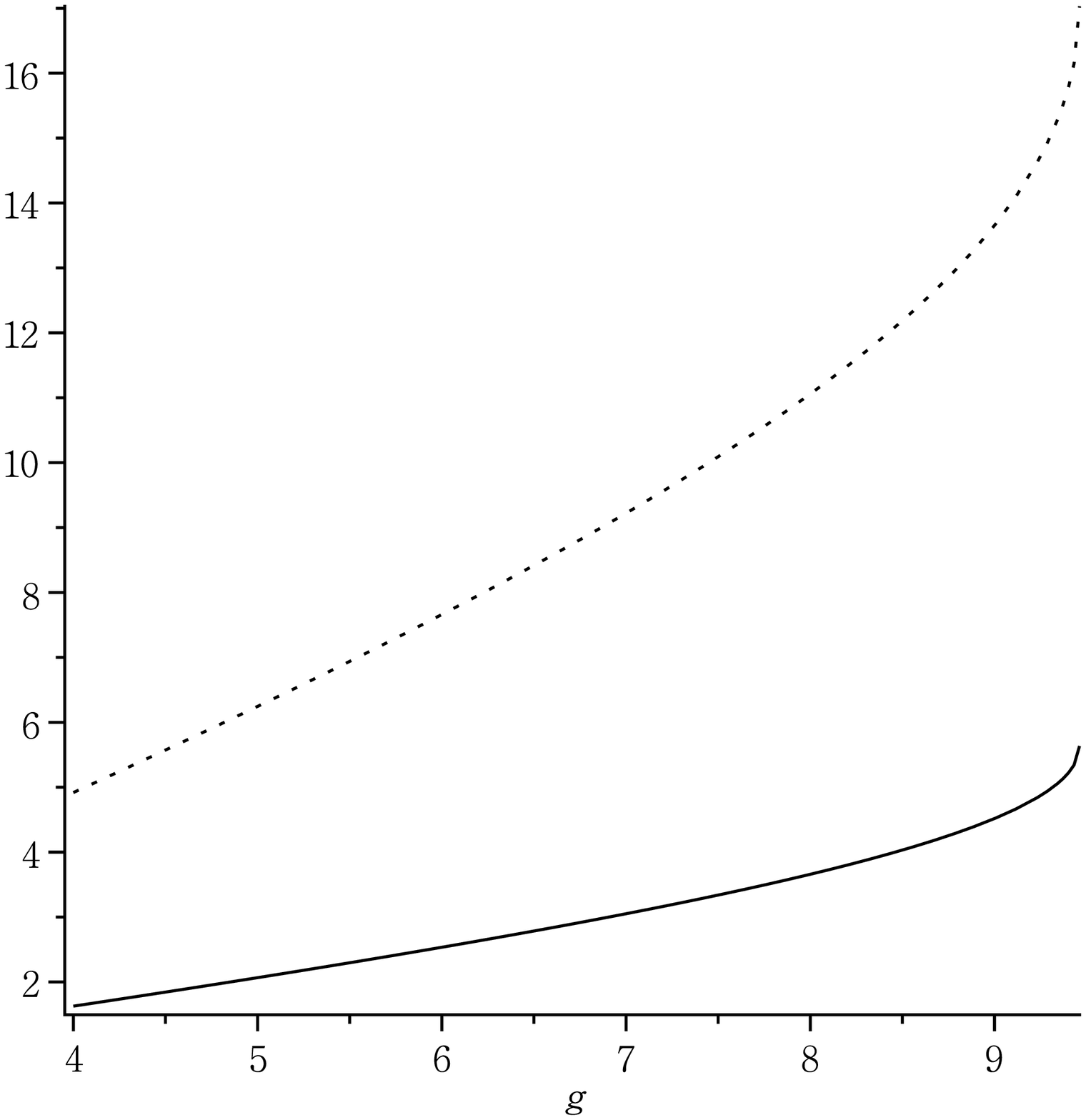}
\end{center}
\caption[]{%
$g_{\rho\pi\pi}^{eff}$ vs g
with the minus
sign in Eq.(\ref{chig}). The lower curve is the
QCD case while the upper curve corresponds to
a hypothetical "walking technicolor" case
with $m_\rho>m_A$.
}
\label{vplot9}
\end{figure}

     In Fig.\ref{vplot9}, which corresponds to the choice of
the minus sign in Eq.(\ref{chig}), it is seen that the
QCD case (lower curve) has no consistent parameter solution
since $g_{\rho\pi\pi}^{eff}$ =8.56 can not be achieved for
$g<9.46$. On the other hand, the upper curve, which corresponds
again to $\tau$ = 1.2, gives reasonable values of
$g_{\rho\pi\pi}^{eff}$.

    To summarize: the QCD case corresponds to the
plus sign choice in Eq.(\ref{chig}) while a possible
consistent parameter solution
 in a non-QCD setting with $m_\rho>m_A$
 is likely to correspond to the minus sign choice.

     It is amusing to observe that the relation between
the vacuum value $v$ and ${\tilde F}_\pi$ differs for the
two sign choices:
\begin{eqnarray}
\label{vandF}
  {\tilde F}_\pi& <& v \hspace{2cm}(+ sign),
\nonumber \\
{\tilde F}_\pi & >& v \hspace{2cm}(- sign).
\end{eqnarray}
To see this note that for the plus sign case, Eq.(\ref{chig})
gives $1/2<\chi<1$ which, using Eq.(\ref{w2}) translates to
$w>\sqrt{2}$ and the desired result when
Eq.(\ref{renor2}) is noted.
The minus sign case is obtained similarly after first
noting $1<w<\sqrt{2}$ in that situation.

    We have seen that the choice of sign in
Eq.(\ref{chig}) distinguishes the two cases where
$m_\rho$ is less than or greater than $m_A$. This 
choice occurs in fitting the parameters to experiment.
It may be of some interest to ask how this distinction is
related to the parameters of the effective Lagrangian
directly. To investigate this, we just subtract the
second of Eqs.(\ref{spinonemass}) from the first:
\begin{equation}
m_\rho^2-m_A^2= -\frac{v^2}{2}(B+g^2).
\label{V-A}
\end{equation}
In the QCD case, Table \ref{params} shows that
$B$ is negative and that the right hand side above
is negative because $g^2>|B|$. In the case which
should correspond to a walking technicolor theory
we evidently must require, if $B$ is also negative, the 
opposite condition $g^2<|B|$. That condition seems
intuitively plausible. Since $B$ is the coefficient
of a scale invariant term in the effective Lagrangian,
we might expect it not to change sign in going from one 
theory to the other. Furthermore, we would expect the
phenomenological coupling constant $g$ to behave something
like the underlying gauge theory coupling constant and hence
to decrease in strength for a ``walking" theory \cite{gwalks}. 

\section{summary and further discussion}

     First, we argued that a detailed treatment of the
gauged minimal linear sigma model would be very helpful
for developing a better understanding of the light
scalar mesons in QCD.

   One quickly realizes, however, that
such a model is much more complicated 
than the ungauged version.
Hence, attention was first paid to developing an 
``analytic procedure" for relating the Lagrangian
parameters to the four well established 
experimental inputs: 
 ${\tilde F}_\pi$, $m_\rho$, $\Gamma_\rho$ and
$m[a_1(1260)]$. 
The key
equation obtained is Eq.(\ref{2signs}) or equivalently, 
Eq.(\ref{chig}).
If the minus sign in this equation is chosen, it 
was shown (in the last section) that there is no consistent
solution of parameters when inputs are taken from the
possible application to QCD of this model. On the other hand,
the minus sign choice allows a solution with $m_\rho>m_A$,
which is plausibly related to a ``minimal walking technicolor"
application of the effective Lagrangian. If the plus 
sign choice is made in this equation, it was shown that the
QCD application of the model is allowed though a 
``walking technicolor" application,
 while possible, seems to correspond
to an extremely large rho boson width. There is 
also a special
case in this equation when the square root vanishes so the
sign choice is irrelevant. In that situation, the Weinberg and 
KSRF relations are both satisfied in the unphysical 
limit where B=0.

    In this work we did not impose either the Weinberg 
or KSRF relations. The model is self-contained
and represents a spontaneously broken chiral SU(2) symmetry
with an assumed particle spectrum of pi, sigma, rho and 
$a_1(1260)$.

    As a check, before comparing the computed s wave pion pion
scattering to experiment, we verified (in the Appendix)
that the complicated formula for the amplitude reduces to that
of the non-linear sigma model without vectors
 when the sigma mass goes to infinity.

    To begin the study of the scattering amplitude in the 
resonant s-wave channel we fit the near threshold NA48/2 data 
\cite{Na48} up to about 370 MeV using the tree amplitude. A good
fit 
was obtained by choosing the bare sigma mass, $m_\sigma$ to
be about 420 MeV. A similarly good fit in the sigma model without
spin 1 fields needed  $m_\sigma$ to be about 620 MeV instead (See
Fig. \ref{vplot3}).
     Once a value of $m_\sigma$ is chosen, the amplitude is
 also predicted
at higher energies. It was pointed out
 (see Fig. \ref{vplot4} that those values of $m_\sigma$
resulted in  ``global" pictures of the s-wave scattering 
which was considerably distorted.  Much better ``global" 
pictures emerge from choices of bare sigma mass, $m_\sigma$
about 850 MeV. However such a value for $m_\sigma$ results in,
as seen in Fig. \ref{vplot10}, some loss of precision for
the region just near threshold. From the standpoint of learning 
about the sigma, the higher bare sigma mass is evidently the 
more suitable one. 

    The feature of getting similar fits with or without vectors,
but with different values of the sigma parameters which emerged
in the discussions in sections V and VI had been 
already observed some time ago \cite{HSS2}. In that case, the 
non-linear 3-flavor chiral Lagrangian was used instead and the region
from threshold to about 1.2 GeV was fit. Rather than the K-matrix
method, a phenomenological unitarization scheme was employed.
 It seems that 
the light
sigma and the $f_0(980)$ are, not surprisingly, the main features of the
I=0, s-wave pion-pion scattering amplitude in this energy range. Adding 
the rho meson
changes somewhat the parameters of the sigma needed for fitting.

    Comparing the ``global" fits to the resonant s-wave pion pion
 scattering amplitude up to about 800 MeV, it is seen that the 
linear sigma model without the spin 1 particles actually gives a
better fit than the one with the spin 1 particles included. This
seems to be due to the higher polynomial terms induced by the
Yang Mills interaction.
    
   If one takes the present model at face value, the
underlying Lagrangian appears to describe all the
contained particles as quark anti-quark composites.
In the mixing picture mentioned earlier \cite{BFS3}-
\cite{ythb}, where all the particles are mixtures
of these ``2 quark" states with ``4 quark" states,
this work should be modified to include two different  
chiral multiplets and the 2-flavor case upgraded to
the 3 flavor case for more realism. This has been done
for the case without vectors and axial vectors in the
mentioned references so adding the spin
 1 particles in that framework is 
a next step. We would expect that the
general behaviors of the lowest lying 
sigma and the rho and $a_1(1260)$ would be similar to 
those in the present model.

    Also in the application to the minimal walking 
technicolor model \cite{wt}, the present piece
 would have to be embedded in a larger framework
 with an initial SU(4) symmetry. One might similarly
expect that the behaviors of the sigma (=Higgs) and the
technicolor spin 1 bosons would be similar to those seen
here \cite{fojasa}.

\section*{Acknowledgments} \vskip -.5cm

We are happy to thank
 A. Abdel-Rehim, D. Black, M. Harada, R. Jora,
S. Moussa, S. Nasri and F. Sannino
 for helpful related discussions.
The work of A.H.F. has been partially
supported by the NSF Grant 0652853.
The work of N.W.P. is
supported by Chonnam National University.
The work of
 J.S. and M.N.S. is supported in part by the U.
S. DOE under Contract no. DE-FG-02-85ER 40231.

\appendix
\section{large sigma mass limit}
We verify here that, in the large sigma mass limit, our
scattering
amplitude reproduces the well-known result
 of the chiral model:
$A(s,t,u)=2(s-{\tilde m}_\pi^2)/{\tilde F}_\pi^2$.

The most crucial piece is the sigma pole contribution
in the s channel given by Eq.(\ref{sigamp}). We start by
expanding the denominator:
\begin{eqnarray}\label{denom}
\frac{1}{m_\sigma^2 -s}=\frac{1}
{m_\sigma^2 -{\tilde m}_\pi^2/w^2}(1+
\frac{s-{\tilde m}_\pi^2/w^2}
{m_\sigma^2 -{\tilde m}_\pi^2/w^2}+
{\cal O}(1/m^4_\sigma)).
\end{eqnarray}
Next we rewrite the numerator of Eq.(\ref{sigamp}) so
as to display the $m_\sigma$ dependence:
\begin{equation}
\frac{w^4}{v^2}(m_\sigma^2-{\tilde m}_\pi^2/w^2)+
-2\frac{w^2}{v}(\sqrt2 g b w {\tilde m}_\pi^2-
2G({\tilde m}_\pi^2-\frac{s}{2}))+\cdots,
\label{num}
\end{equation}
where the dots indicate terms
 independent of $m_\sigma^2$.
These two equations give the result for Eq.(\ref{sigamp}):
\begin{equation}
\frac{w^4}{v^2}(m_\sigma^2-{\tilde m}_\pi^2/w^2)+
\frac{w^4}{v^2}(s-{\tilde m}_\pi^2/w^2)
-2\frac{w^2}{v}(\sqrt2 g b w {\tilde m}_\pi^2-
2G({\tilde m}_\pi^2-\frac{s}{2})).
\label{NonD}
\end{equation}
Notice that the first term above is cancelled
by Eq.(\ref{amp1}). For simplicity we will consider
the C=0 case since the C contribution ends up
not contributing in the present limit. Then
Eq.(\ref{2dercon}) just gives:
\begin{equation}
-Bb^2w^2{\tilde m}_\pi^2+b^2w^2(B+\frac{g^2}{2})s.
\label{2der}
\end{equation}
Next, note that the contribution of Eq.(\ref{4dercon})
is higher order in ${\tilde m}_\pi^2$ and hence
 negligible for the present purpose. Finally,
the rho pole contribution of Eq.(\ref{rhoamp})
 to leading order has the form
\begin{equation}
G_1^2\frac{3s-4{\tilde m}_\pi^2}{m_\rho^2}.
\label{rho}
\end{equation}
The net result so far consists of Eq.(\ref{NonD})
 without the first term,
Eq.(\ref{2der}) and Eq.(\ref{rho}).

We use the following expressions to simplify these terms:

\begin{eqnarray}\label{af1}
2\sqrt2\frac{w^2}{v} g b w = \frac{4}{v^2}w^2(w^2-1),
\nonumber\\
4G\frac{w^2}{v} = \frac{4}{v^2}(w^2-1)(w^2+1)-
\frac{2B}{g^2v^2}(w^2-1)^2,\nonumber\\
G_1^2 = \frac{g^2}{2}(1-\frac{Bv^2}{2{m^\prime}_0^2})^2,
\nonumber\\
b^2w^2= \frac{2}{g^2v^2}(w^2-1)^2.
\end{eqnarray}

 Then, the part of $A(s,t,u)$ proportional to $s$ is:

\begin{eqnarray}\label{as}
s[\frac{1}{v^2}((3-2w^2)+\frac{3B}{g^2}(w^2-1)^2+
\frac{3g^2v^2}{2m_\rho^2}(1-\frac{B}{g^2}(w^2-1))^2)],
\end{eqnarray}

while the part proportional to ${\tilde m}_\pi^2$  is:

\begin{eqnarray}\label{ampi}
 {\tilde m}_\pi^2[-\frac{w^2}{v^2}-\frac{4}{v^2}w^2(w^2-1)+
\frac{4}{v^2}(w^2-1)(w^2+1)-\frac{4B}{g^2v^2}
(w^2-1)^2- \frac{2g^2}{m_\rho^2}(1-\frac{B}{g^2}(w^2-1))^2].
\end{eqnarray}

 We will now show that Eq.(\ref{as}) is equal to
$\frac{2s}{{\tilde F}_\pi^2}$. For this,
 we notice that the terms which
do not include B in Eq.(\ref{as})  become

 \begin{eqnarray}\label{as1}
s[3-2w^2+\frac{3g^2v^2}{2m_\rho^2}]
=s[3-2w^2+\frac{3}{2m_\rho^2}(2m_\rho^2+Bv^2)(w^2-1)]
=s[w^2+\frac{3Bv^2}{2m_\rho^2}(w^2-1)],
\end{eqnarray}

where we used,
\begin{equation}\label{arel}
g^2v^2=2{m^\prime}_0^2(w^2-1)=(2m_\rho^2+Bv^2)(w^2-1).
\end{equation}

Then, all B terms in Eq.(\ref{as}) and in
Eq.(\ref{as1}) can be arranged to give:

\begin{eqnarray}\label{aB1}
s[\frac{3B}{g^2}(w^2-1)(\frac{g^2v^2}
{2m_\rho^2}-w^2+1-\frac{Bv^2}{2m_\rho^2}(w^2-1))].
\end{eqnarray}

It is easily shown that this is equal to zero
using the relation (\ref{arel}).
Then, we can see that Eq.(\ref{as}) becomes
$\frac{sw^2}{v^2}=\frac{2s}{{\tilde F}_\pi^2}$,
 which is the desired result.

The ${\tilde m}_\pi^2$ part can be treated similarly.
  The terms which
do not include B in Eq.(\ref{ampi}) become

\begin{eqnarray}\label{ampi1}
 {\tilde m}_\pi^2[
-\frac{w^2}{v^2}-\frac{4}{v^2}w^2(w^2-1)
+\frac{4}{v^2}(w^2-1)(w^2+1)-\frac{2g^2}{m_\rho^2}]\nonumber\\
= {\tilde m}_\pi^2[-\frac{w^2}{v^2}-\frac{2B}{m_\rho^2}(w^2-1)].
\end{eqnarray}
The first term on the second line,
$-{\tilde m}_\pi^2 w^2/v^2$ is just $-2{\tilde m}_\pi^2/{\tilde
F}_\pi^2$, while the second term is proportional to B. Then
 all terms linear in B in Eq.(\ref{ampi}) become

\begin{eqnarray}\label{ampi2}
{\tilde m}_\pi^2[-\frac{2B}{m_\rho^2}(w^2-1)
-\frac{4B}{g^2v^2}(w^2-1)+\frac{4B}{m_\rho^2}(w^2-1)]\nonumber\\
={\tilde m}_\pi^2[\frac{2B}{g^2v^2}(w^2-1)
\Big[2(w^2-1)-\frac{g^2v^2}{m_\rho^2}\Big]]
={\tilde m}_\pi^2[\frac{2B^2}{g^2m_\rho^2}(w^2-1)^2],
\end{eqnarray}

where in the second line, we used the relation,
 Eq.(\ref{arel}). This term is exactly
the negative of the  $B^2$ term in Eq.(\ref{ampi}),
 so all B terms
cancel out to give the desired result.


\begin{thebibliography}{10}

\bibitem{gl}M. Gell-Mann and M. Levy, Nuovo Cimento, {\bf 16},
705(1960). Here, we consider the version of this model 
where the nucleon fields are absent.

\bibitem{njl}See also Y. Nambu and G. Jona-Lasinio,
Phys. Rev. {\bf 122}, 345 (1961), {\bf 124}, 246 (1961).

\bibitem{ad}S.L. Adler and R.F. Dashen, ``Current Algebras
and Application to Particle Physics", W.A. Benjamin, New York,
1968.

\bibitem{ws}S. Weinberg, Phys. Rev. Lett. {\bf 19}, 1264 (1967);
A. Salam, p. 367 of {\it Elementary Particle Theory}, ed.
N. Svartholm (Almquist and Wiksells, Stockholm, 1969). 

\bibitem{wt}A detailed review of the "minimal walking technicolor"
model is given in R. Foadi, M.T. Frandsen, T.A. Ryttov
 and F. Sannino, Phys. Rev. D {bf 76}, 055005 (2007).
See also F. Sannino, arXiv:0804.0182[hep-ph].
 An historical
review of the subject is given by K. Yamawaki, Proc. of the
 International Symposium pn$\Lambda$ 50, PTP supplement no. 167, 127
(2007).

\bibitem{kyotoconf}See the dedicated conference proceedings, S.
Ishida et al
``Possible existence of the sigma meson and its implication to hadron
physics", KEK Proceedings 2000-4, Soryyushiron Kenkyu 102, No. 5, 2001.
Additional points of view are expressed in the proceedings, D. Amelin
and A.M. Zaitsev ``Hadron Spectroscopy'', Ninth International Conference
on Hadron Spectroscopy, Protvino, Russia(2001) and A. H. Fariborz,
``Scalar mesons, an interesting puzzle for QCD", Utica N. Y. (2003),
AIP Conference Proceedings Vol. 688.

\bibitem{vanBev} E. van Beveren, T.A. Rijken, K. Metzger,
C. Dullemond, G. Rupp and J.E. Ribeiro, Z. Phys. {\bf C30}, 615
(1986). E. van Beveren and G. Rupp, hep-ph/9806246, 248.  See also
J.J. de Swart, P.M.M. Maessen and T.A. Rijken, U.S./Japan Seminar on the
YN Interaction, Maui, 1993 [Nijmegen report THEF-NYM 9403].

\bibitem{MP}
D. Morgan and M. Pennington, Phys. Rev. {\bf D48},  1185  (1993).

\bibitem{BMPV} A.A. Bolokhov, A.N. Manashov, M.V. Polyakov and
V.V. Vereshagin, Phys. Rev. {\bf D48}, 3090 (1993).  See also
V.A. Andrianov and A.N. Manashov, Mod. Phys. Lett. {\bf A8}, 2199
(1993).  Extension of this string-like approach to the $\pi K$ case
has been made in V.V. Vereshagin, Phys. Rev. {\bf D55}, 5349 (1997)
and  in A.V. Vereshagin and V.V. Vereshagin, {\it{ibid.}} {\bf
59}, 016002 (1999).

\bibitem{AS94} N.N. Achasov and G.N. Shestakov, Phys. Rev. {\bf
D49}, 5779 (1994).

\bibitem{Kam94}{R. Kam\'inski}, {L. Le\'sniak} and J. P. Maillet,
Phys. Rev. {\bf D50}, 3145 (1994).

\bibitem{SS}F.~Sannino and J.~Schechter, Phys. Rev.  {\bf D52},  96
(1995).

\bibitem{T}{N.A.~T\"ornqvist}, Z. Phys.
{\bf C68}, 647 (1995) and references therein.  In addition see
{N.A.~T\"ornqvist} and M. Roos, Phys. Rev. Lett. {\bf 76}, 1575
(1996), N.A. T\"ornqvist, hep-ph/9711483 and Phys. Lett. {\bf B426} 105
(1998).

\bibitem{DS} R. Delbourgo and M.D. Scadron, Mod. Phys. Lett. {\bf
A10}, 251 (1995).  See also D. Atkinson, M. Harada and A.I. Sanda,
Phys.~Rev. {\bf D46}, 3884 (1992).

\bibitem{JPHS}
{G.~Janssen, B.C.~Pearce, K.~Holinde and J.~Speth}, Phys. Rev. {\bf
D52},  2690  (1995).

\bibitem{Sv} M. Svec, Phys. Rev. {\bf D53}, 2343 (1996).

\bibitem{Ishida} S. Ishida, M.Y. Ishida, H. Takahashi, T. Ishida,
K. Takamatsu and T Tsuru, Prog. Theor. Phys. {\bf 95}, 745 (1996),
S.~Ishida, M.~Ishida, T.~Ishida, K.~Takamatsu and T.~Tsuru,
Prog. Theor. Phys. {\bf 98}, 621 (1997). See also M. Ishida and S. Ishida,
Talk given at 7th International Conference on Hadron Spectroscopy (Hadron
97), Upton, NY, 25-30 Aug. 1997, hep-ph/9712231.

\bibitem{HSS1}M. Harada, F. Sannino and J. Schechter, Phys. Rev. {\bf
D54},
1991 (1996).

\bibitem{HSS2}M. Harada, F. Sannino and J. Schechter,
Phys. Rev. Lett. {\bf 78}, 1603 (1997).

\bibitem{BFSS1}D. Black, A.H. Fariborz, F. Sannino and J. Schechter,
Phys. Rev. {\bf D58}, 054012 (1998).

\bibitem{BFSS2}D. Black, A.H. Fariborz, F. Sannino and J. Schechter,
 Phys. Rev. {\bf D59}, 074026 (1999).
 
\bibitem{OOP} J.A. Oller, E. Oset and J.R. Pelaez, Phys. Rev. Lett.
{\bf 80}, 3452 (1998). See also K. Igi and K. Hikasa, Phys. Rev. {\bf
D59}, 034005 (1999).

\bibitem{AnSa}A.V. Anisovich and A.V. Sarantsev, Phys. Lett. {\bf B413},
137 (1997).

\bibitem{EFSS}V. Elias, A.H. Fariborz, Fang Shi and T.G. Steele,
Nucl. Phys. {\bf A633}, 279 (1998).

\bibitem{Dm} V. Dmitrasinovi\'c, Phys. Rev. {\bf C53}, 1383 (1996).

\bibitem{MO}P. Minkowski and W. Ochs, Eur. Phys. J. {\bf C9}, 283 (1999).

\bibitem{GN}S. Godfrey and J. Napolitano, hep-ph/9811410.

\bibitem{BG}L. Burakovsky and T. Goldman, Phys. Rev. {\bf D57}
2879 (1998)

\bibitem{Hatsuda}T. Hatsuda, T. Kunihiro and H. Shimizu, Phys. Rev. Lett.
 {\bf 82}, 2840 (1999); S. Chiku and T. Hatsuda,
 Phys. Rev. D {\bf 58}, 076001 (1998).

\bibitem{Shakin}L. Celenza, S-f Gao, B. Huang and C.M. Shakin,
Phys. Rev. C {\rm 61}, 035201 (2000).

\bibitem{j}A diquark, anti-diquark nonet was
explored in:
R.L.~Jaffe,
Phys.\ Rev.\ D {\bf 15}, 267 (1977).
A partial meson-meson nonet was advocated in:
J.D.~Weinstein and N.~Isgur,
Phys.\ Rev.\ Lett.\  {\bf 48}, 659 (1982) and
also,
 Y.~S.~Kalashnikova, A.~E.~Kudryavtsev,
 A.~V.~Nefediev, C.~Hanhart and
J.~Haidenbauer,
  Eur.\ Phys.\ J.\ A {\bf 24} (2005) 437.

\bibitem{roy}The Roy equation for the pion amplitude, S.M. Roy, Phys.
Lett. B {\bf 36}, 353 (1971), has been used by several authors to obtain
information about the $f_0(600)$ resonance. See T. Sawada, page 67 of
ref. \cite{kyotoconf} above, I. Caprini, G. Colangelo and H. Leutwyler,
Phys. Rev. Lett. {\bf 96}, 132001 (2006). A similar approach has been
employed to study the putative light kappa by S.Descotes-Genon and
B. Moussallam, Eur. Phys. J. C {\bf 48}, 553 (2006).

\bibitem{b}Further discussion of the approach in ref. \cite{roy} above
is given in D.V. Bugg, J. Phys. G {\bf 34}, 151 (2007) [hep-ph/0608081].
\bibitem{zhs} A. Zhang, T. Huang and T.G. Steele, hep-ph/0612146.

\bibitem{BFS3}D. Black, A. H. Fariborz and J. Schechter,
Phys. Rev. {\bf D61} 074001 (2000).

\bibitem{BFMNS01}D. Black,
A.H. Fariborz, S. Moussa, S. Nasri and J. Schechter,
 Phys.Rev. {\bf D64}, 014031 (2001). See also
A.H. Fariborz, R. Jora and J. Schechter, Phys. Rev. D
 {\bf 77}, 094004 (2008) and E. Meggiolaro, Z. Phys. C
 {\bf 62}, 669 (1994).

\bibitem{mixing}
 T. Teshima, I. Kitamura and N. Morisita, J. Phys. G
{\bf 28}, 1391 (2002); {\it ibid} {\bf 30}, 663 (2004); F. Close and N.
Tornqvist, {\it ibid.}
{\bf 28}, R249 (2002); A.H. Fariborz, Int. J. Mod. Phys. A {\bf 19},
2095 (2004); 5417 (2004); Phys. Rev. D {\bf 74}, 054030 (2006);
F. Giacosa, Th. Gutsche, V.E. Lyubovitskij and A. Faessler,
Phys. Lett. B {\bf 622}, 277 (2005); J. Vijande, A. Valcarce,
F. Fernandez and B. Silvestre-Brac, Phys. Rev. D {\bf 72}, 034025
 (2005); S. Narison, Phys. Rev. D {\bf 73}, 114024 (2006); L. Maiani,
 F. Piccinini, A.D. Polosa and V. Riquer, hep-ph/0604018.
 J.R. Pelaez,
Phys. Rev. Lett. {\bf 92}, 102001 (2004); J.R. Pelaez and G. Rios,
Phys. Rev. Lett. {\bf 97}, 242002 (2006); F. Giacosa,
Phys. Rev. D {\bf 75},054007 (2007).

\bibitem{NR04}M. Napsuciale and S. Rodriguez, Phys. Rev. D {\bf 70},
 094043 (2004).

\bibitem{FJS05}A.H. Fariborz, R. Jora and J. Schechter,
Phys. Rev. D {\bf 72}, 034001 (2005);
Phys. Rev. D {\bf 76}, 114001 (2007).

\bibitem{thooft} G. 't Hooft, G. Isidori, L. Maiani,
A.D. Polosa and V. Riquer, Phys. Lett. B {\bf 662}, 424 
(2008).

\bibitem{liu}In K-F Liu, arXiv:0706.1262 [hep-ph], the author
presents evidence from lattice theory for a picture of a scalar
spectrum containing light four quark type states and heavier
two quark type states.

\bibitem{ythb}In N. Yamamoto, M. Tachibana, T. Hatsuda and G. Baym,
 arXiv:0704.2654 [hep-ph]
and A.A. Andrianov and D. Espriu, arXiv:0709.0049
[hep-ph]
 similar models are
discussed for non zero temperature and pressure.



\bibitem{gg}S. Gasiorowicz and D.A. Geffen, Rev. Mod.
Phys. {\bf 41}, 531 (1968).

\bibitem{recentsigmod}
M.D. Scadron,
F. Kleefeld and G. Rupp,
arXiv:hep-ph/0601196; N. Achasov,
 arXiv:0810.2201[hep-ph];
S. Struber and D.H. Rischke, Phys. Rev. D 
{\bf 77}, 085004 (2008);
D. Parganlija, F. Giacosa and D.H. Rischke,
arXiv:0812.2183[hep-ph].

\bibitem{fran}T. Applequist and F. Sannino, Phys. Rev. D 
{\bf 59}, 067702 (1999); R. Foadi, M.T. Frandsen, T.A. 
Ryttov and F. Sannino, Phys. Rev. D {\bf 76}, 055005 (2007).

\bibitem{Weinberg}S. Weinberg, Phys. Rev. Lett. {\bf 18}, 
507 (1967); T. Das, V. Mathur and S. Okubo, Phys. Rev. Lett.
{\bf 18}, 761 (1967).

\bibitem{KSRF}K. Kawarabayashi and M. Suzuki, Phys. Rev. 
Lett. {\bf 16}, 255 (1966); Riazuddin and Fayazuddin, Phys. 
Rev. {\bf 147}, 1071 (1966).

\bibitem{car}S. Weinberg, Phys. Rev. Lett. {\bf 17}, 616
 (1966).

\bibitem{Na48}J.R. Batley et al [NA48/2 Collaboration],
Eur. Phys. J. C {\bf 52}, 875 (2007)[arXiv:0707.0697].
See also
 R. Kaminski, J.R. Pelaez and F.J. Yndurian,
Phys. Rev. D {\bf 77}, 054015 (2008).

\bibitem{nochange}See the discussion
 around Eq.(47)
in the second reference of \cite{FJS05} given above.

\bibitem{kggp}R. Kaminski, R. Garcia-Martin, P. Grynkiewicz
and J.R. Pelaez, arXiv:0811.4510[hep-ph].

\bibitem{gwalks} See for example Fig. 5 in the second
paper in \cite{wt} above.

\bibitem{fojasa}See also R. Foadi, M. Jarvinen and 
F. Sannino, arXiv:0811.3718[hep-ph].

\end{thebibliography}
\end{document}